\documentclass[pra,reprint,aps,longbibliography]{revtex4-1}
\usepackage[latin1]{inputenc}
\usepackage{graphicx}
\usepackage{subfigure}
\graphicspath{{./bilder/}}
\usepackage{amsmath}
\usepackage{amsthm}
\usepackage{amssymb}
\usepackage{amsfonts}
\usepackage{bbm}
\usepackage[verbose]{placeins}
\usepackage{color}
\usepackage{times}

\usepackage{natbib}
\usepackage[colorlinks,bookmarks=false,citecolor=blue,linkcolor=red,urlcolor=blue]{hyperref}

\DeclareMathOperator{\Tr}{Tr}
\newcommand{\vect}[1]{\boldsymbol{#1}}

\begin{document}


\title{Chiral Mott insulators in frustrated Bose-Hubbard models on ladders and two-dimensional lattices:\\ a combined perturbative and density matrix renormalization group study}
\author{Christian Romen}
\author{Andreas M. L\"auchli}
\affiliation{Institut f\"ur Theoretische Physik, Universit\"at Innsbruck, A-6020 Innsbruck, Austria}
\date{\today}

\begin{abstract}
We study the fully gapped chiral Mott insulator (CMI) of frustrated Bose-Hubbard models on ladders and two-dimensional lattices by perturbative strong-coupling analysis and density matrix renormalization group (DMRG). First we show the existence of a low-lying exciton state on all geometries carrying the correct quantum numbers responsible for the condensation of excitons and formation of the CMI in the intermediate interaction regime. Then we perform systematic DMRG simulations on several two-leg ladder systems with $\pi$-flux and carefully characterize the two quantum phase transitions. We discuss the possibility to extend the generally very small CMI window by including repulsive nearest-neighbour interactions or changing density and coupling ratios.
\end{abstract}

\maketitle

Ultracold atoms in optical lattices are a promising approach to study exciting many-body phenomena appearing in condensed matter systems, but also provide interesting
many body platforms in their own right.
The first experimental realization \cite{BoseHubbardGreiner} of one of the most simple interacting bosonic models, the Bose-Hubbard model \cite{BoseHubbardFisher},
showed that a quantum system can be preparated and manipulated in a controlled way such that phase transitions triggered by pure quantum fluctuations can be 
observed in the lab and hence reveal different states of matter.
This motivated theorists and experimentalists to study strongly-interacting models and seek novel phases.
By using synthetic gauge fields \cite{syntgauge2009JakschZoller,syntgauge2011,syntgauge2014Lewenstein,HarperHamiltonian,HofstadterRealization} one can create
spin-orbit coupling \cite{spinOrbit1} and artificial magnetic fields, which lead to rich magnetic lattice physics such as the fractal Hofstadter-butterfly \cite{Hofstadter}, 
topological Chern insulators with chiral edge states \cite{syntheticHallRibons2015,edgestatesHallRegime}, 
chiral spin superfluids, quantum Hall and spin Hall 
states \cite{Li2014,chiralLaddersQuantumHallHuegel,arxiv_EdgeModesTaddia}. 
Using Raman transitions effective magnetic fields \cite{syntgauge2009} and spin-orbit coupling \cite{spinOrbitExp1,spinOrbitExp2,spinOrbitExp3} were generated successfully in 
experiments. 
An alternative technique for generating synthetic gauge fields is time-dependent shaking of optical potentials \cite{timedependentShakingExp,Struck2013}.

Recently, the observation of chiral currents in bosonic ladders \cite{Atala2014} drew intensified attention to ladder systems.
The interplay of interactions, anisotropic couplings, filling and flux reveals many different phases 
(an overview is given in Ref.~\cite{PhysRevA.94.063628}).
For strong fields the translation symmetry can be broken by forming a vortex lattice crystal with variable vortex 
density $\rho_{V}$ \cite{greschner_reversalCurrent,PhysRevA.94.063628}, so called vortex lattice phases or chiral phases emerge.
If the magnetic flux is small a chiral current is flowing only around the border screening the external magnetic field \cite{MeissnerEffectGiamarchi} similar as the
 Meissner-Ochsenfeld effect in superconductors.
Meissner and vortex phases in bosonic two-leg ladders have been discussed for hardcore bosons
 \cite{meissnerVortexHardcoreDiDio,piraud_hardcoreBosons,Orignac_hardcoreBosons,crepin_rungMottHardcoreBosons,Orignac_Incommensurate_phases} 
as well as softcore bosons 
\cite{greschner_reversalCurrent,keles_artificialMagField,artificialMagFieldFieldTheoretical,Petrescu_Le_Hur_meissner_currents,Tokuno_population_imbalance,PhysRevA.94.063628}.
 The fully frustrated case at $\pi$-flux
 and unit-filling 
contains a small chiral Mott insulator phase, which is a fully gapped vortex crystal with $\rho_v=1/2$ \cite{Paramekanti2012,Paramekanti2013}.
Frustrated hopping can also be achieved without gauge fields on a triangular lattice by changing the hopping sign \cite{zalatel2014}.
In quasi-1D, bosons on the zig-zag ladder without and with an additional three-body constraint can show charge density waves, bond ordered insulators, 
chiral superfluidity, chiral Mott insulators, pair superfluidity and even a chiral Haldane-Insulator phase \cite{PhysRevB.87.174504,greschner_zigzag}.
Meissner and vortex-lattice phases naturally arise as well on three-leg ladder systems with homogeneous flux \cite{three_leg_ladder}.

\begin{figure}[h!]
\centering 
\includegraphics[width=0.40\textwidth]{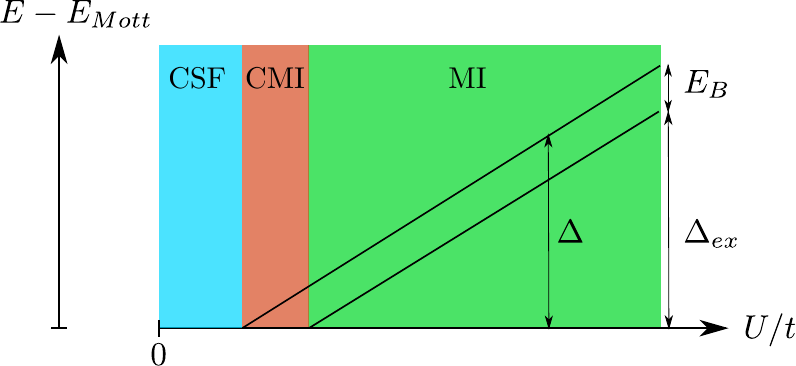}
\caption{Qualitative schematic representation of the excitation gap $\Delta_{\text{ex}}$ and particle-hole gap $\Delta$ for the CSF-CMI-MI scenario.}
\label{fig:exciton_state}
\end{figure}

The goal of the present study is to deepen our understanding of the puzzlingly small chiral Mott insular phase observed in Refs. \cite{Paramekanti2012,Paramekanti2013} and to find ways to
enlarge the chiral Mott insulating phase (first results in this direction have appeared in Ref.\cite{zalatel2014}). Our approach is two-fold. On the one hand we perform
a perturbative strong coupling analysis of the structure of bound states in the Mott insulating phase of the Bose-Hubbard model in one and two dimensions in
the presence of uniform orbital magnetic flux $\pi$ per plaquette. We find that in all geometries there is a formation of a particle-hole bound state (an exciton)
with the correct quantum numbers to lead to a chiral Mott insulating phase in case this bound state condenses - without concomitant single particle condensation -
as function of the interaction. This scenario is illustrated in Fig.~\ref{fig:exciton_state}. We find that the bound states are more weakly bound in two space dimensions
compared to one-dimensional systems. This might point to a enhanced fragility for the prospects of finding chiral Mott insulating phases in two-dimensional frustrated Bose-Hubbard models. 
After the strong coupling perspective we move to a systematic density matrix renormalization group (DMRG) study of several two-leg ladder-like systems at unit filling and 
we particularly study the behaviour of the exciton gap from strong to intermediate coupling, and investigate the effect of an additional nearest-neighbour density repulsion
on the extent of a chiral Mott insulator phase. As the CMI phases remain small in extent we study a two-leg ladder at density per site of 1/2, where we find a much larger 
CMI phase. We then proceed to a detailed investigation of the two phase transitions delimiting the CMI phase. Surprisingly we find that the Ising transition between the CMI
and Mott insulator is quite challenging to characterize, despite the small central charge.

\begin{figure}[h!]
\centering 
\includegraphics[width=0.35\textwidth]{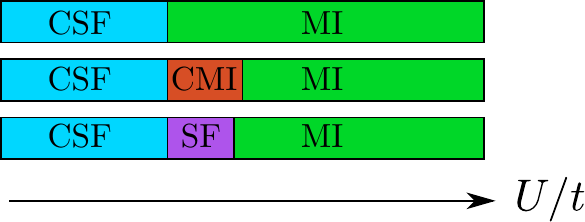}
\caption{Three possible scenarios for the fully frustrated Bose-Hubbard phase diagram on the triangular lattice as a function of on-site interaction.
Possible phases are ordinary superfluid (SF), chiral superfluid (CSF), chiral Mott insulator (CMI) and the ordinary Mott insulator (MI). }
\label{fig:phasediagram}
\end{figure}

\begin{figure}[h!]
\centering 
\includegraphics[width=0.35\textwidth]{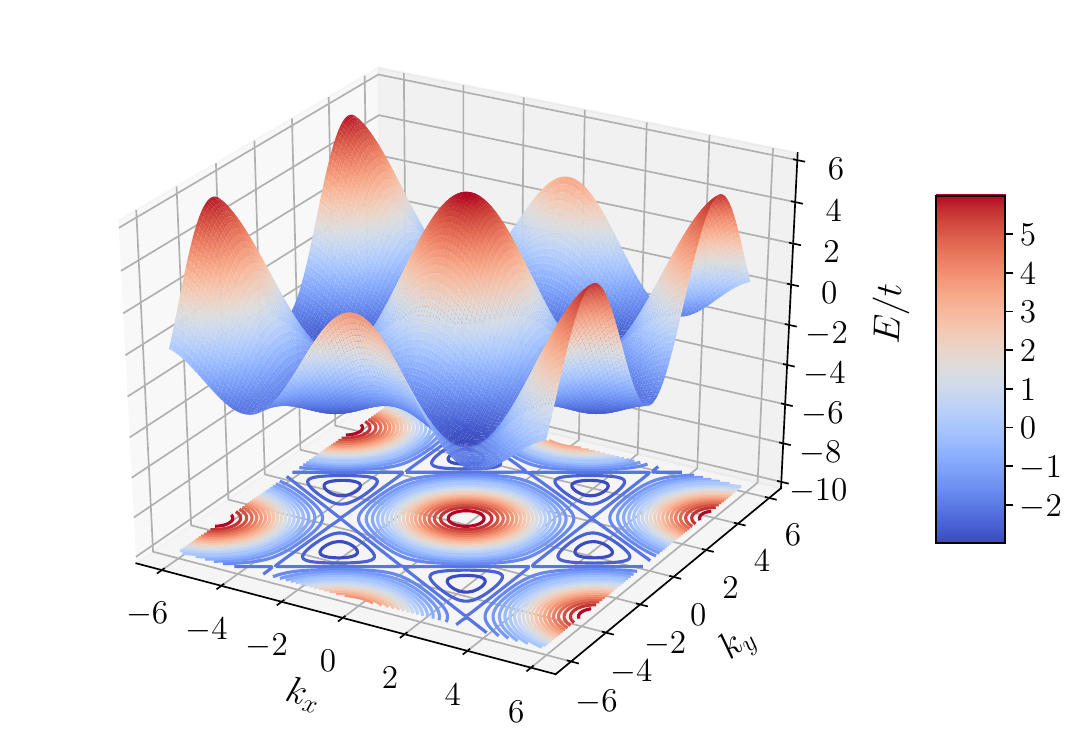}
\caption{Single particle dispersion in units of the tunneling amplitude $t$ on the triangular lattice.}
\label{fig:trilatt_dispersion}
\end{figure}

We start with a Bose-Hubbard model on the triangular lattice with unity-filling
\begin{equation}
  H = t \sum_{\left<ij\right>} \left( b_i^\dagger b_j + \text{h.c.} \right) + \frac{U}{2} \sum_i n_i^2,
\end{equation}
with the usual bosonic creation (annihilation) operators $b_i^\dagger$ ($b_i$), positive on-site repulsion energy $U$ and inverted isotropic hopping amplitude $t$.
 Considering the non-interacting case $U=0$, discrete Fourier-transform yields the single particle dispersion 
$\epsilon(\vect{k}) = 2t\left[ \cos (k_x) + 2 \cos (k_x/2) \cos (\sqrt{3}k_y/2) \right] $ illustrated in Fig.~\ref{fig:trilatt_dispersion}. 
Instead of the unique minimum at the origin $K=\Gamma$ 
in the unfrustrated case the dispersion has two nonequivalent minima at $K=(4\pi/3,0)$ and $K'=(2\pi/3,2\pi/\sqrt{3})$.
That means that the non-interacting ground state is highly degenerate since each boson can condense at either $K$ or $K'$.
For small $U$, the ground state within mean-field theory \cite{Paramekanti2012,Paramekanti2013} consists of a equal superposition between
 $\mathbf{k}=K$ and $\mathbf{k}=K'$ superfluid states with two possible relative phases corresponding to a $Z_2$ symmetry which leads to a 
staggered current pattern in real-space breaking the $C_6$ and time-reversal symmetry in addition to the $U(1)$ symmetry breaking associated
to the superfluid nature.
 The loop currents around plaquettes can be seen as vortices and form an antiferromagnetic vortex-antivortex crystal.
This phase is called \textit{chiral} superfluid.
For large $U\gg t$ the system pays too much energy penalty by occupying only two modes, and individual bosons get pinned to the sites and 
a large charge gap is present.  A transition to a Mott insulator without currents and with a uniform density has occurred~\cite{zalatel2014,Paramekanti2012,Paramekanti2013}. 
How precisely this transition happens at intermediate $U/t$ is an interesting question. 

Generically, three abstract scenarios are possible (see Fig.~\ref{fig:phasediagram}) within a Ginzburg-Landau picture: chirality is lost at the same time as the superfluidity,
 i.e. only one phase transition exists, which is either first order or (fine-tuned) second order, where 
$U(1)$, time-reversal and the $C_6$ symmetry are restored at once. 
The second possibility is to have two second order transitions defining a finite window where superfluidity is lost
but still maintains chirality with staggered loop currents, the fully gapped chiral Mott insulator (CMI).
The third scenario is having two second order phase transitions, where chirality is lost in a first step, when coming from weak interactions,
obtaining a ordinary superfluid. With the second transition a charge gap opens and $U(1)$ symmetry is restored.

In order to discuss which of the scenarios is realized in the investigated Bose-Hubbard models, we define the following gap quantities.
We start by defining the {\em neutral} excitation gap
\begin{equation}
 \Delta_{\mathrm{ex}}=E_1(N)-E_0(N)
\label{eq_gap_ex}
\end{equation}
as the energy difference between the ground state and the first excited state, where in both cases the sector with $N$ bosons 
in the finite-size system of a given size is targeted. 
Further we define the particle-hole gap 
\begin{equation}
\begin{split}
 \Delta &= [E_0(N+1)-E_0(N)]+[E_0(N-1)-E_0(N)] \\
	&= E_0(N+1)+E_0(N-1)-2E_0(N)
\end{split}
\label{eq_gap_delta}
\end{equation}
as the energy of (independently) adding and removing one particle to the Mott ground state. It is precisely this gap which goes to zero if 
superfluidity or Bose-condensation is taking place.
Since the energy of adding one particle or one hole is sometimes called the single particle gap, the 
particle-hole gap defined here (\ref{eq_gap_delta}) should be understood as twice the single particle gap.
It measures the energy gap to the threshold of the particle-hole scattering continuum.
An interesting binding occurs when the binding energy defined as
\begin{equation}
 E_B = \Delta_{\mathrm{ex}} - \Delta
 \label{eq_bindingenergy}
\end{equation}
is negative, i.e.~when the particle-hole gap is larger than the excitation gap.

Coming from the strong coupling regime, the CMI phase in the second scenario can be seen as a condensate of bound particle-hole pairs (excitons)~\cite{HalperinRice}.
In this picture an exciton, lying lower than any other neutral excitation, exists within the first excited particle-hole Hubbard band region and is separated from the Mott ground state
by the gap $\Delta_{\text{ex}}$, illustrated in Fig.~\ref{fig:exciton_state}. The particle-hole gap is larger than the excitation gap, i.e. $E_B < 0$. 
With shrinking interaction $U$ the gap $\Delta_{\mathrm{ex}}$ vanishes first and a condensate of 
excitons starts to form. If $\Delta$ is nonzero at this point a finite window exists where the system is gapped but supports staggered loop currents 
around the elementary plaquettes. 
The chiral superfluid at weak interaction and the Mott insulator at large $U$ can be described and understood within mean-field theory \cite{Paramekanti2013}, 
which is not the case for the chiral Mott insulator. It can only appear in the intermediate regime $U\sim t$ making it challenging to detect and characterize.

The question we want to tackle is how can we understand microscopically the chiral Mott insulator in a perturbative way. The region amenable to us is the strong coupling regime.

This paper is structured as follows: In the first section we derive an effective Hamiltonian in the particle-hole subspace on the triangular lattice with perturbation
theory. We go up to second order in $t/U$ and show that a low-lying excitonic bound state exists, suitable for the CMI phase. We consider different geometries and include also an additional repulsive nearest-neighbour interaction \cite{zalatel2014}. In the second section we perform finite-size DMRG simulations on a two-leg ladder with $\pi$-flux
 and a zig-zag ladder, where we switch on a nearest-neighbour interaction for each. We confirm the existence 
of a very small CMI phase. We conclude by analyzing the phase transitions delimiting a sizeable chiral {\em rung} Mott insulator at half-filling for the anisotropic case $t_\perp/t=2$.

\section{\label{lab:perturbation}Exciton aspect in strong coupling regime}

\subsection{First order}

In this section we study the strong-coupling regime with perturbation theory and try to find exciton states which are important for the CMI in the intermediate regime.
In the strong-coupling regime $U\gg t$ at unity-filling the Hubbard spectrum is split up into different branches, the Hubbard bands. 
The lowest band consists of only one state, the uniformly 
filled Mott state. Higher bands correspond to subspaces with one or more multi occupancies. We are considering only the second band with 
one double occupation (doublon) and one vacancy (holon) on top of a uniformly filled state. In lowest order perturbation theory (first order in $t/U$) the effective Hamiltonian 
is a simple projection to our considered subspace

\begin{equation}
 H_{\mathrm{eff}} = PVP + PH_0P,
\label{eq_effHamil_firstOrder}
\end{equation}
where $V$ is the perturbation (hopping term) and $H_0$ the interaction term. Since we have one double occupation in our subspace, $PH_0P$ yields clearly $U$, which is only a
global shift of the spectrum. $PVP$ results in a simple hopping process of doublon and holon
 \begin{equation}
  PVP \left|21\right>=2t\left|12\right>, \; PVP \left|01\right>=t \left|10\right>, \; PVP\left|11\right> = 0
 \end{equation}
where the hopping amplitude for the doublon is twice as large due to the Bose factor. The uniformly filled Mott state is not affected by this order.
 In other words, in lowest order the effective Hamiltonian is a two-body problem
with the restriction that doublon and holon cannot sit on the same site. Formally written

\begin{equation}
 H_{\mathrm{eff}} = \sum_{\left< i,j \right>} [2t \, d_i^\dagger (1-h_i^\dagger h_i)d_j + \mbox{h.c.}+t \, h_i^\dagger (1-d_i^\dagger d_i)h_j + \mbox{h.c.}]
\label{eq:firstOrderH}
\end{equation}
with new creation (annihilation) operators $d_i^\dagger$, $h_i^\dagger$ ($d_i$, $h_i$) for doublon and holon, respectively. This Hamiltonian can be easily solved
numerically with Exact Diagonalization (ED).
To characterize bound states in the following we use two quantities: first, the binding energy defined in equation (\ref{eq_bindingenergy}),
 $E_B = \Delta_{\mathrm{ex}}-\Delta$. Second, the binding length measuring the spatial extension of the wavefunction, 
$\xi^2 := \sum_{\vect{r}_{\mathrm{rel}}} \left| \psi(\vect{r}_{\mathrm{rel}}) \right|^2 \cdot r^2 $, where $\psi(\vect{r}_{\mathrm{rel}})$ denotes the probability amplitude of the holon if 
the doublon is fixed at the origin and $r=\left| \mathbf{r}_{\mathrm{rel}} \right|$ the distance betwen doublon and holon. If doublon and holon form a bound state the wavefunction is spatially localized and
therefore $\xi$ does ultimately not increase with the system size.
Within first order $\Delta$ can be evaluated easily. The reference Mott energy is zero $E_0(N)=0$ and the minimum of the free holon dispersion is $-3t$ located at 
the $K$ and $K'$ points.
The doublon has twice the hopping amplitude and hence the neutral particle-hole gap is $\Delta=U+\epsilon_d(K)+\epsilon_h(K)=U-9t$.

\begin{figure}[h!]
\centering 
\includegraphics[width=0.49\textwidth]{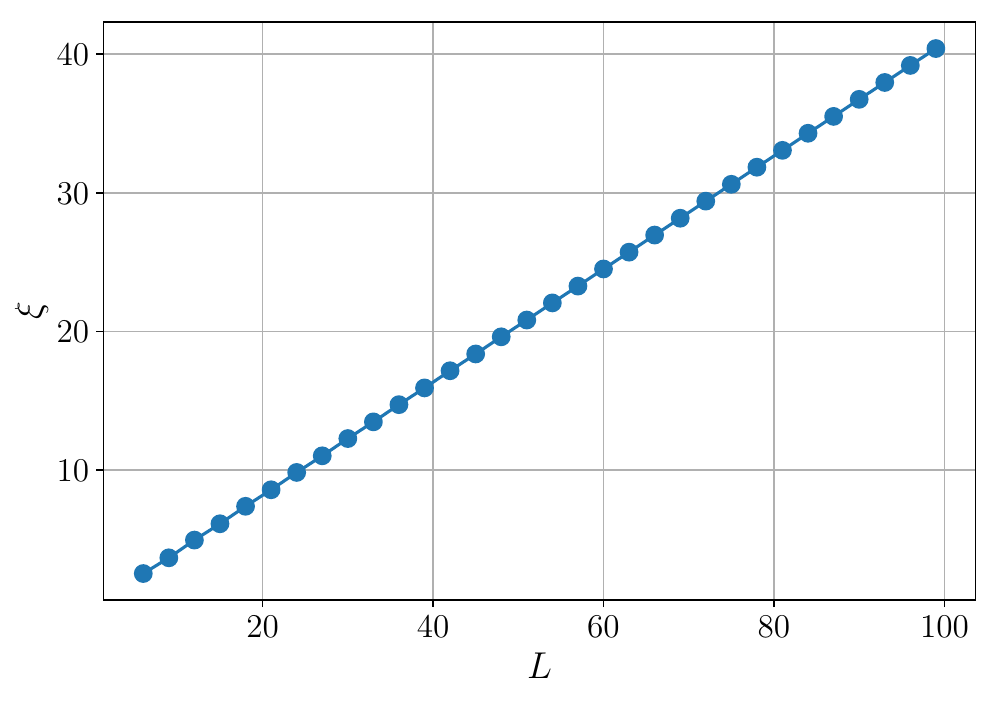}
\caption{
Ground state binding length as a function of system size of 
the model (\ref{eq:firstOrderH}) on the triangular lattice.
The perfect linear behavior suggests a free hopping of holon and doublon.}
\label{fig:firstOrder_xi}
\end{figure}

Now we solve Eq. (\ref{eq:firstOrderH}) numerically.
In fact, Fig.~\ref{fig:firstOrder_xi}(a) shows a linear behavior of $\xi$ with the linear system size $L_x$, giving evidence for a free doublon and holon and exhibited by 
a wavefunction uniformly spread over the whole lattice. Interestingly, by fixing the doublon on one site (e.g. the origin in our case) and plotting the relative wavefunction 
we notice that zero probability points appear forming a regular pattern.
Starting from such a knot, by translating $3 \vect{a}_1$ or $\vect{a}_1+\vect{a}_2$ the next knot is reached, which produces itself a triangular lattice on top. 
Hence, the hardcore constraint leads to destructive interference and extensively many lattice sites are forbidden for the holon. By calculating the complex 
phase of the relative wavefunction on every point we can infer that the ground state is described by the irreducible representation $B1$ of the $C_{6v}$ group,
 which is antisymmetric with respect to a rotation of $\pi/3$, $\pi$ and antisymmetric with respect to a reflection through the lattice bonds. 
 It carries zero momentum and has an energy $E=U-9t$ for all system sizes, therefore the binding energy is exactly zero, $E_B = 0$.
The three lowest excitations are $K A1$ (doubly degenerate) and $\Gamma A1$. For all three states the relative wave function is almost zero around the origin and 
increases with the radius, which means doublon and holon want to maximize the 
distance to each other, we call this antibinding. With increasing system size all states come closer together and form a gapless excitation spectrum in the thermodynamic limit.
The first important result: \textbf{Up to first order in $t/U$ no exciton exists in the low-energy sector}. 

\subsection{Additional nearest-neighbour interaction}

\begin{figure}[h!]
\centering 
\includegraphics[width=0.49\textwidth]{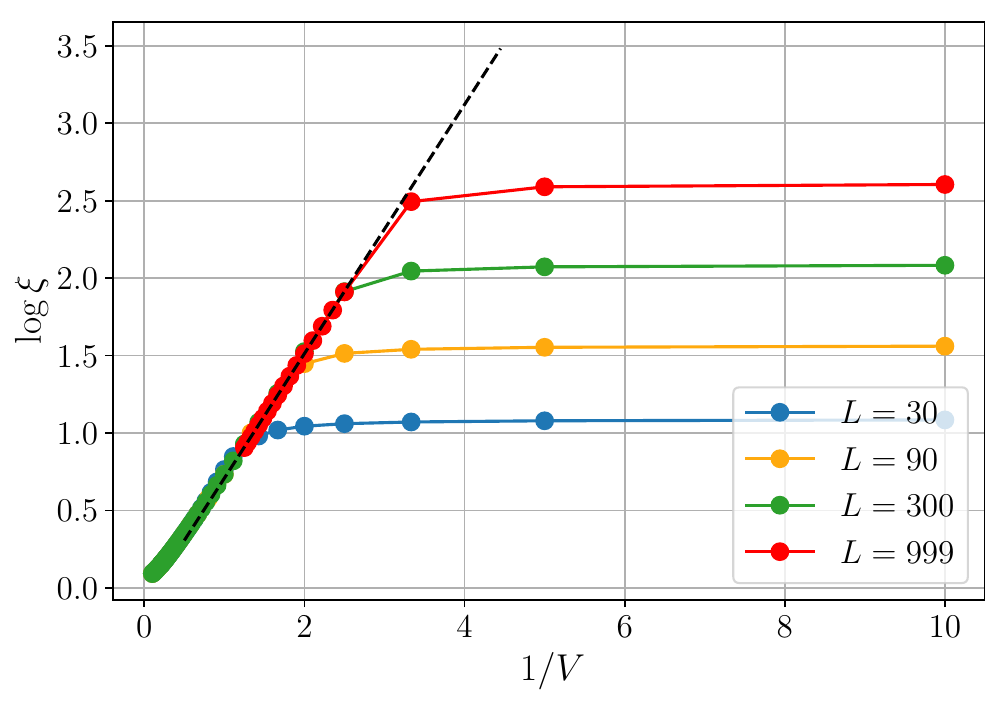}
\includegraphics[width=0.49\textwidth]{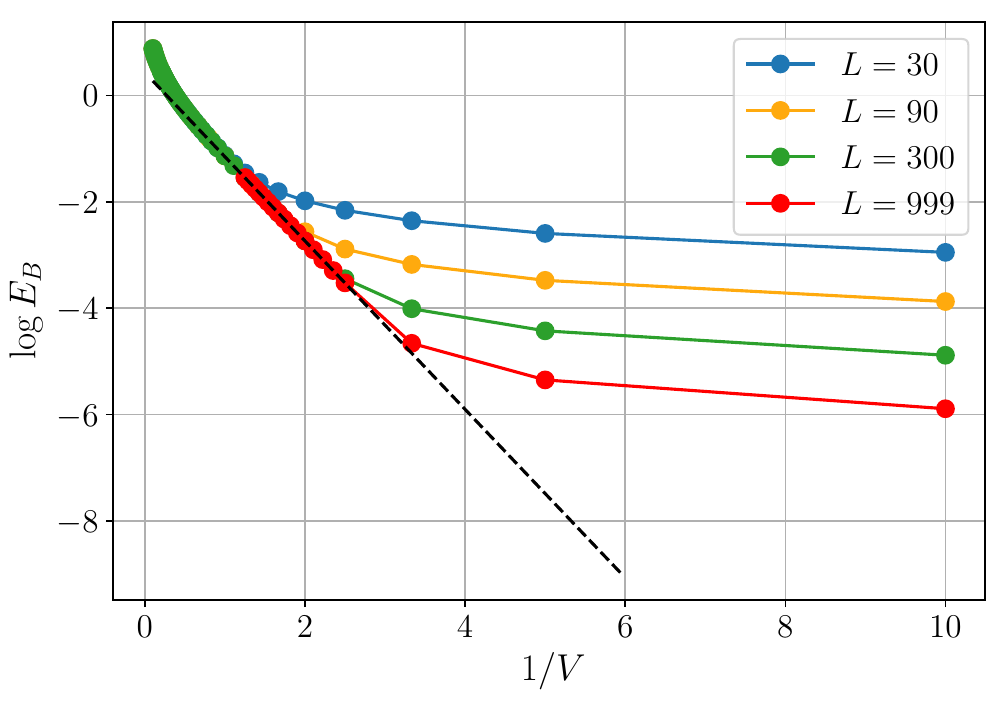}
\caption{ED ground state results of model (\ref{eq:firstOrderH}) on the triangular lattice with additional nearest neighbour interaction $V$. 
Logarithm of binding length and binding energy as a function of the inverse interaction $1/V$. The dashed line is a linear fit to all system sizes.}
\label{fig:firstOrder_xiEB}
\end{figure}

Before going to the second order we study the effect of adding a repulsive nearest neighbour interaction as suggested in Ref. \cite{zalatel2014}. 
In the original Hubbard language the term has the form 
$V \sum_{\left< ij \right>} (n_i-n_0) (n_j-n_0)$ with $n_0=1$ for unity-filling. In the particle-hole subspace all matrix elements of this term are zero except if
 doublon and holon are neighbours, which gives $-V$, i.e. an attractive interaction arises. To quantify this effect we include the term and redo the ED calculation of the
ground state. Fig.~\ref{fig:firstOrder_xiEB} shows the logarithm of the binding length (top) and binding energy (bottom) as a function of the inverse interaction.

One can nicely see the data collapse for large enough interactions $V$. Between $1 \lesssim 1/V \lesssim 2.5$ both quantities exhibit linear behavior
 signalized by the dashed black line. The linear regime would extend to arbitrarily large $1/V$, i.e. small $V$ if even larger system sizes would be used, hence the 
drifting away from linearity at small $V$ is a finite-size effect.
  Linear behavior on a logarithmic y-axis means exponential on the linear scale
\begin{align}
 \log |E_B| &\propto -\alpha/V \Rightarrow E_B \propto -e^{-\alpha/V} \\
\log \xi   &\propto \beta/V \Rightarrow \xi \propto e^{\beta/V}. \nonumber
\end{align}
It turns out that $\alpha \approx 2\beta$ and hence $E_B \propto -\xi^{-2}$. In the strong interaction regime $V\gg 1$, it is intuitive that $V$ is the leading scale 
and thus $E_B \propto -V$ and $\xi \propto V$, which can be seen by plotting $E_B$ and $\xi$ on a double logarithmic scale (not shown). 

Physically the interpretation is that finite interaction creates an attraction between the two particles, leading to a bound state, localized in space. 
If the characteristic length $\xi$ of the wavefunction is much smaller than the linear system size $L$, nothing changes by increasing $L$, corresponding to a 
data collapse. By decreasing $V$ the wavefunction gets more delocalized, which leads to a decrease of the binding and increase of $\xi$.
 If $\xi$ has roughly the same order of magnitude as $L$, the wavefunction
``feels'' the border and gets dependent of the system size. Therefore, in order to verify a shallow bound state in the thermodynamic limit, the system size has to be huge. 

The ground state wavefunction shows the same symmetry as before ($\Gamma B1$) but lowers its energy with respect to 
the particle-hole state. Even in the thermodynamic limit a finite gap dependent on $V$ separates the ground state from the rest of the spectrum.
 The ground state wavefunction is still modulated with the knots on the sublattice.
The rest of the low-energy spectrum does not change qualitatively, a doubly degenerate $K A1$ and $\Gamma A1$, both antibinding. 
Other bound states occur at higher energies
and lie in the quasi-continuum. Increasing $V$ leads to more than one bound state in the low-energy sector.

Summarized, by turning on an arbitrarily weak repulsive interaction $V$ the binding energy opens exponentially in the $2D$ system.

\subsection{Ladder case}

\begin{figure}[h!]
\centering 
\includegraphics[width=0.49\textwidth]{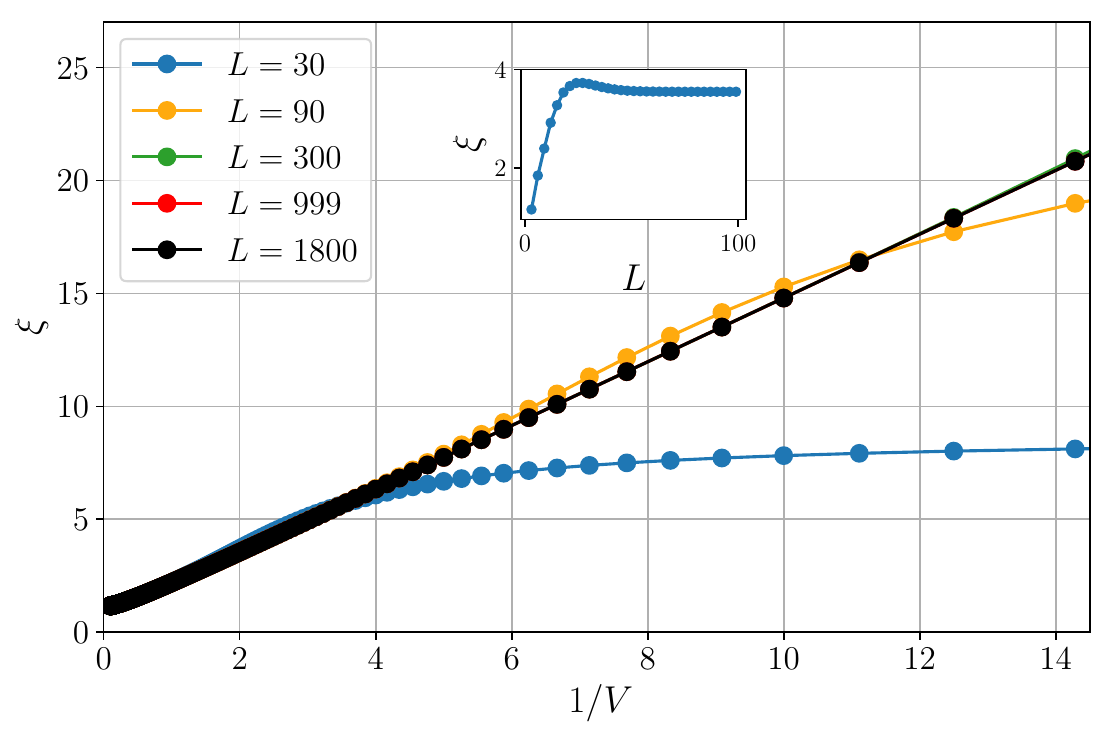}
\includegraphics[width=0.49\textwidth]{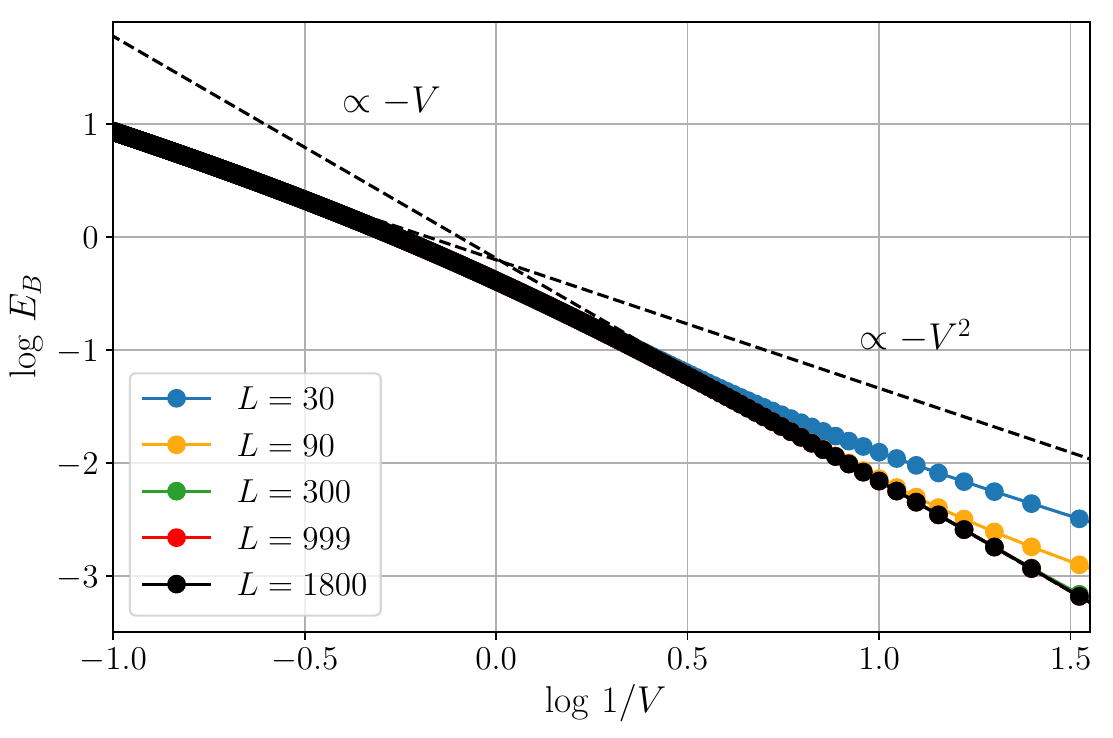}
\caption{Ground state results of model (\ref{eq:firstOrderH}) on the triangular lattice
with additional repulsive nearest-neighbour interaction $V$ in the asymmetric case $L_y=3$, $L_x\rightarrow\infty$.
 (top) Binding length as a function of the inverse interaction. The inset shows the finite-size
behavior of the binding length for $V=0.5$.
 (bottom) Logharithm of the binding energy as a function of 
the logarithm of the inverse interaction. The dashed lines are fits in the two limiting cases $V\ll 1$ and $V\gg 1$ with the slopes $2$ and $1$, respectively.}
\label{fig:firstOrder_asymp_xiEB}
\end{figure}

In this subsection we investigate the situation in one-dimension by fixing $L_y=3$ and sending $L_x\rightarrow\infty$. In the following we denote $L=L_x$. 
Fig.~\ref{fig:firstOrder_asymp_xiEB} (top) shows $\xi$ as a function of $1/V$. Qualitatively, compared to the $2D$ case both curves look the 
same (though for $2D$ the y-axis was $\log \xi$) except that the bending down is smoother and 
happens at smaller $V$ for a given system size. In that way the two largest system sizes merge in the whole region down to $V=0.1$. 
For small $V$, the binding length behaves linearly, 
$\xi \propto 1/V$, in contrast to the exponential behavior in $2D$. 
The inset shows the finite-size behavior of $\xi$ for $V=0.5$. If $L\gg \xi$ the binding length starts to saturate. Surprisingly, 
$\xi$ slightly decreases before going to saturation.
Fig \ref{fig:firstOrder_asymp_xiEB} (bottom) shows $\log E_B$ as a function of $\log 1/V$. 
One can nicely see the crossover regime at $V \sim 1$ which separates the strong interaction regime with $E_B \propto -V$ from the weak interaction regime with
a quadratic increase of the binding energy, $E_B \propto -V^2$. This result reflects the strong dependence of the dimensionality which is reminiscent of the different 
scaling behaviors of the binding energy in basic studies of single particle quantum mechanics \cite{boundstates_scaling}, i.e. in $1D$ the binding energy opens 
quadratically, in $2D$ exponentially and in higher dimensions a finite binding threshold exists. 

\subsection{Second order}
\begin{figure}[h!]
\centering 
\includegraphics[width=0.20\textwidth]{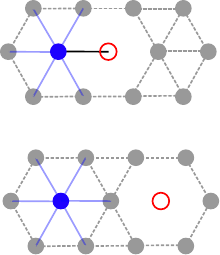}
\caption{Pictorial representation for the formation of static binding in second order. Boson-boson bonds are dashed black lines, doublon-boson bonds solid blue lines
and the doublon-holon bond is represented as a solid black line. The two possible types of configurations are: 
doublon (filled blue circle) and holon (empty red circe) are neighbours (top) or separated (bottom). }
\label{fig:secOrder_statBind}
\end{figure}

In the previous subsections we considered within first order perturbation theory an additional $V$ term and the one dimensional limit. Now 
we go back to the two dimensional case and include second order terms.
The general expression for the effective second order Hamiltonian is \cite{Takahashi_perturbation}
\begin{equation}
  H_{\text{eff}} = PVP + PH_0P+PVQ\frac{1}{U-H_0}QVP,
\label{eq:perturb_sec_order}
\end{equation}
where $Q=1-P$ ensures that the subspace is temporarily left such that the denominator never vanishes.
By analyzing the action on different basis states we identify the following second order processes:
\begin{itemize}
 \item next-nearest neighbour hopping of each particle
      \\ $\left|211\right> \rightarrow -2t^2/U \left|112\right>$,
      \\ $\left|011\right> \rightarrow -2t^2/U \left|110\right>$
 \item shift (collective hopping)
      \\ $\left|201\right> \rightarrow 2t^2/U \left|120\right>$
 \item exchange 
      \\ $\left|20\right> \rightarrow 2t^2/U \left|02\right>$
 \item tunneling 
     \\ $\left|201\right> \rightarrow 2t^2/U \left|102\right>$, 
     \\ $\left|021\right> \rightarrow t^2/(2U) \left|201\right>$,
\end{itemize}
and the diagonal terms
\begin{align*}
 \left|20\right> &\rightarrow 2t^2/U \left|20\right>, \hspace{0.82cm} \left|11\right> \rightarrow -4t^2/U \left|11\right> \\
 \left|21\right> &\rightarrow -3t^2/(2U) \left|21\right>, \; \left|01\right> \rightarrow 0 \left|01\right>,
\end{align*}
where the last term is blocked. Embedding the diagonal terms on a triangular lattice generates a static attraction as illustrated in Fig.~\ref{fig:secOrder_statBind}. 
Doublon-boson bonds are solid blue lines, boson-boson lines are dashed black lines and the doublon-holon bond is marked as a solid black line.
The state where doublon and holon are neighbours is energetically favored by an amplitude of $-t^2/(2U)$, which means an effective $V$ term appears. 

\begin{figure}[h!]
\centering 
\includegraphics[width=0.50\textwidth]{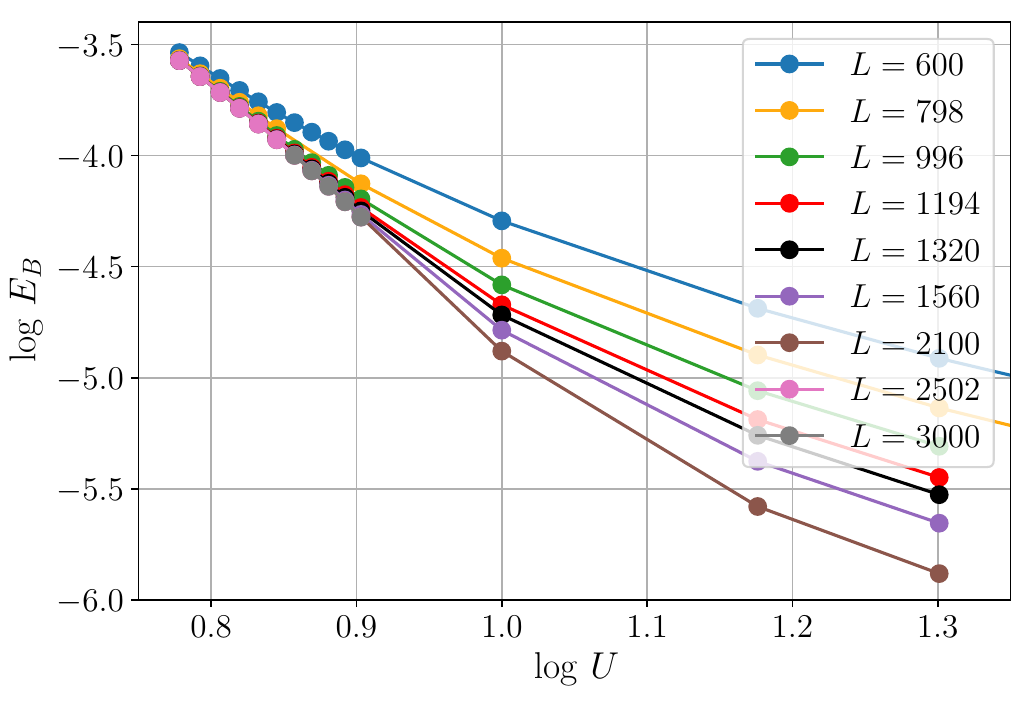}
\includegraphics[width=0.50\textwidth]{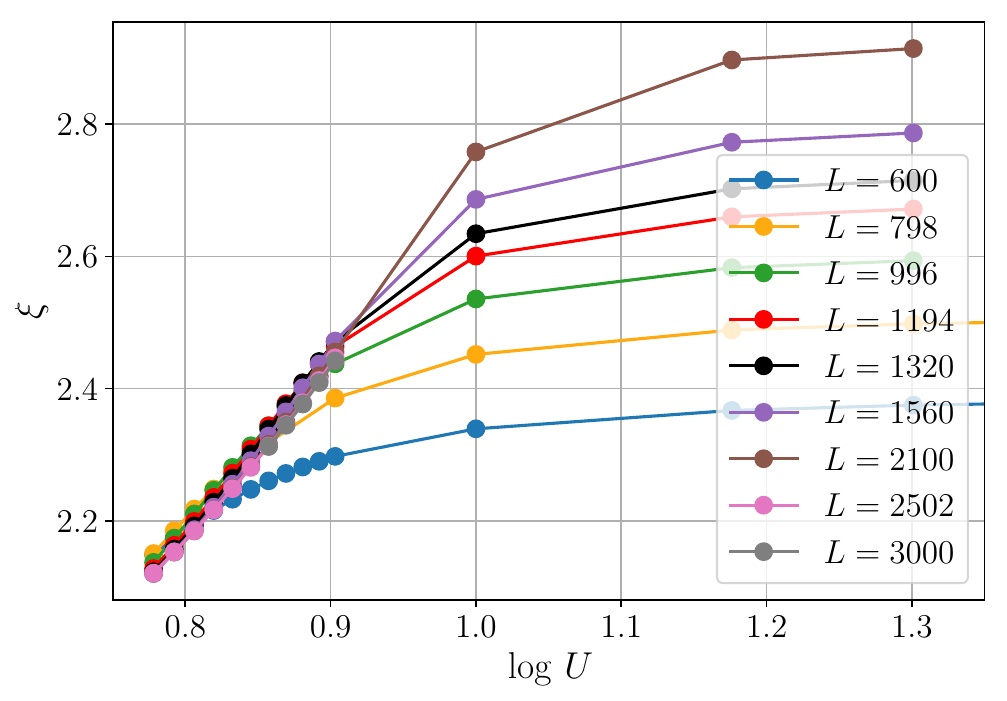}
\caption{Logarithm of the ground state binding energy and binding length as a function of the logarithmic interaction of the 
model (\ref{eq:perturb_sec_order}) on the triangular lattice.}
\label{fig:secOrder_triangl}
\end{figure}

It is not a priori clear which of the diagonal and off-diagonal terms actually lead to the binding of the doublon and the holon. 
Including the processes of Eq. (\ref{eq:perturb_sec_order}) yields the ground state binding energy and binding length shown in Fig.~\ref{fig:secOrder_triangl}. 
Below $U \sim 7$ we are able to make the system sizes large enough to converge the two quantities, though in this regime second order perturbation theory 
does not give a reliable result for the original Hubbard model.
However, the model (\ref{eq:perturb_sec_order}) exhibits a finite but very small binding energy $E_B \sim 10^{-4}$.
The straight line on a double logarithmic scale reveals a power-law behavior of $E_B$ and $\xi$ in $U$.
A calculation with only first order hopping and the diagonal terms of second order shows nearly linearly increasing $\xi$ at $U=6$ which shows 
that the generated glue originates mostly from the {\em off-diagonal} terms.
By including different combinations of second order terms we study the relative strength and notice that the shift term is a single repulsive one. 
All the other terms act attractively. On the triangular lattice the next-nearest neighbour hopping can be decomposed into second order nearest neighbour hopping (\textit{soNN}) 
and pure next-nearest neighbour hopping (\textit{pNNN}) since the smallest loop consists of three sites. Within this decomposition \textit{pNNN} acts repulsively and \textit{soNN}
gives the most dominant contribution to attraction. 

The low-energy sector is formed by the three states $\Gamma B1$, $K A1$ (doubly degenerate) and $\Gamma A1$.
The ground state $\Gamma B1$ is slightly separated from the rest of the spectrum (on the order of the binding energy), which forms a continuum in the thermodynamic limit.
Indeed, the lowest state $\Gamma B1$ is important for the CMI phase. This can be seen as follows: In the CMI staggered loop currents along elementary
 triangles yield two current patterns and form together a doubly degenerate 
ground state. For both patterns the unit cell is composed of two triangles which recur at multiples of the primitive vectors of the lattice, and hence the translation symmetry is not
broken.
If translation symmetry is not spontaneously broken the ground state representation can not contain a nonzero momentum. In the exciton picture,
it can be shown that the current operator creates a doublon and a holon with momenta $K$ and $K'$ in an antisymmetric way and respective zero total momentum,
 analog to the calculation in \cite{Paramekanti2013}.
From a group theoretical point of view it follows that the two-fold degenerate ground space decomposes into two irreducible representations, $\Gamma A1$ and $\Gamma B1$.
Therefore, in the intermediate coupling regime where the particle-hole gap is small but finite and the excitation gap vanishing, the excitonic bound state together with the symmetric Mott state $\Gamma A1$ form a two-fold degenerate ground space, and  allow to break time-reversal and $C_6$ symmetry in the thermodynamic limit.
 
\subsection{Square lattice with $\pi$-flux}

\begin{figure}[h!]
\centering 
\includegraphics[width=0.15\textwidth]{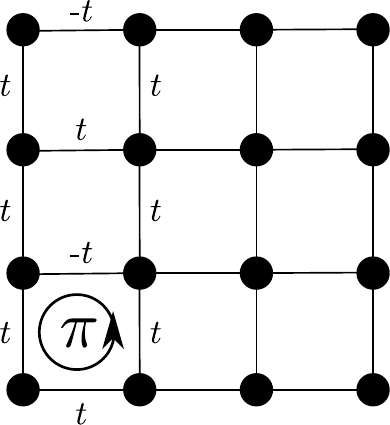}
\caption{Pictorial representation of the square lattice with $\pi$-flux through every plaquette. The flux is achieved by changing the hopping sign on every odd leg.}
\label{fig:squareCouplings}
\end{figure}

In this short section we investigate the perturbative exciton behavior on the square lattice with $\pi$-flux.
Repeating the analysis from the previous chapter we target the first excited state and study its properties within strong coupling expansion up to second order.
In order to obtain frustration one can introduce a gauge field. Generally, by 
using the Peierls substitution \cite{Hofstadter} the hopping 
amplitudes become complex and the phase $e/\hbar \int_{\vect{r}_i}^{\vect{r}_j} d\vect{r} \cdot \vect{A}(\vect{r}) $ is accumulated by hopping from site $i$ to $j$.
For a closed loop the accumulated phase $e/\hbar \oint_C d\vect{r} \cdot \vect{A}(\vect{r}) = 2\pi\Phi/\Phi_0 $ is proportional to the ratio between
 the magnetic flux penetrating the area enclosed by the path $C$, $\Phi$, and the magnetic flux quantum $\Phi_0=h/e$.
By choosing the Landau gauge $\vect{A} = -By\vect{e}_x$, $\vect{B} = B \vect{e}_z$ and the coordinate system such that all $y$-values are integers,
 only horizontal hopping gives an additional factor $e^{i\lambda y}$, where $\lambda$ is the flux per plaquette.

We choose $\pi$-flux, $\lambda=\pi$, by changing the hopping sign on every odd line, as shown in Fig.~\ref{fig:squareCouplings}.
Due to the chosen gauge the unit cell is twice as large in $y$-direction and hence the Brillouine zone is divided into half ($k_y \in [0,\pi)$).
In the noninteracting case a discrete Fourier transform yields the dispersion relation $\epsilon(\vect{k}) = \pm 2t \sqrt{\cos(k_x)^2+\cos(k_y)^2}$ with 
two minima at $\vect{k}=\Gamma$ and $\vect{k}=(\pi,0)$. 

First order perturbation theory $t/U$ yields the two-body problem of Eq. \ref{eq:firstOrderH}.
Asymptotic separation of both particles is related to the gap $\Delta = U + \epsilon_d(0)+\epsilon_h(0)= U-6\sqrt{2}t$.
The ground state of this model is no bound state, visible by a wavefunction spread over the whole lattice. 
As for the triangular lattice there are zero probability points, forming itself a square lattice with lattice vectors $2\vect{a}_1$ and $2\vect{a}_2$ with respect to
the location of the doublon. Finite nearest-neighbour repulsion 
induces an exponentially small binding energy $E_B \propto -e^{-c/V}$. The bound state carries momentum $\vect{k}=(\pi,0)$ and is totally symmetric under point
group operations. 

Embedding second order terms yields two important differences to the triangular case: due to the gauge the next-nearest neighbour hopping
 to the site $\vect{r}+\vect{a}_1+\vect{a}_2$ cancels out for both particles, except if they are neighbours. Second, the nearest neighbour hopping doesn't get
renormalized. Together with the smaller coordination number it leads to a very shallow bound state, one order of magnitude weaker than on the triangular lattice.
But notice that this bound state does not lie in the regime where the second order perturbation theory gives a reliable result. However, 
for the proof of principle we verify that the ground state of the model (\ref{eq:perturb_sec_order}) is a bound state at $U=0.1$ with a 
saturating binding length for $L \ge 800$. 

In principle the bound state fulfills the requirements to form together with 
the trivial Mott state $\Gamma$A1 a degenerate ground state and break the translation invariance and time-reversal symmetry.

\subsection{2-leg ladder with $\pi$-flux}

\begin{figure}[h!]
\centering 
\includegraphics[width=0.25\textwidth]{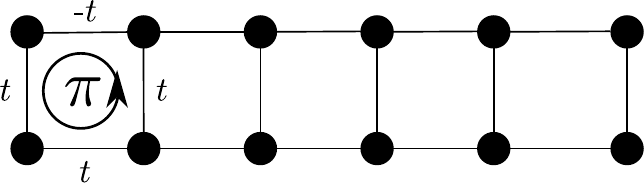}
\caption{Pictorial representation of the two-leg ladder with $\pi$-flux through every plaquette. The flux is achieved by changing the hopping sign on the upper leg.}
\label{fig:ladderCouplings}
\end{figure}

In this section we change the geometry and cover the (quasi) $1d$-case by considering the 2-leg ladder with $\pi$-flux. 
We use periodic boundary conditions in $x$ direction and open boundaries in $y$ direction. Hence, the wave vector is a one-dimensional quantity and takes values
$k=2\pi/L\cdot n$ with $n=0,\dots,L-1$. 
The $\pi$-flux is imposed by choosing the Landau gauge and the coordinate system such that the lower (upper) leg lies at $y=0$ ($y=1$), respectively.
For $\lambda=\pi$, hopping along the upper leg gives a minus sign (shown in Fig.~\ref{fig:ladderCouplings})
 resulting in a accumulated phase of $\pi$ around a plaquette, as on the square lattice.

In the noninteracting case the dispersion relation is $\epsilon(k) = \pm t_{\perp} \sqrt{1+4(t/t_{\perp})^2\cos(k)^2}$ with 
two minima at $k=0$ and $k=\pi$. In the following we consider the isotropic case $t_{\perp}=t$.

\begin{figure}[h!]
\centering 
\includegraphics[width=0.20\textwidth]{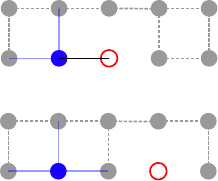}
\caption{Pictorial representation for the formation of static binding in second order on the 2-leg ladder. Boson-boson bonds are dashed black lines, doublon-boson bonds 
solid blue lines and the doublon-holon bond is represented as a solid black line. The two possible types of configurations are: 
doublon (filled blue circle) and holon (empty red circe) are neighbours (top) or separated (bottom). }
\label{fig:secOrder_statBindLadder}
\end{figure}

In first order $t/U$ perturbation theory we obtain again the two-body Hamiltonian (\ref{eq:firstOrderH}). 
From the dispersion we get the particle-hole gap $\Delta = U+\epsilon_d(0)+\epsilon_h(0)= U-3\sqrt{5}t$.
Numerical diagonalization reveals the same 
result as for the $2d$ lattices, an exactly non-binding ground state, (with momentum $k=\pi$) $E_B=0$ with $\xi\propto L$.
It is striking that the exact non-binding of model (\ref{eq:firstOrderH}) seems to be largely independent of the geometry and dimensionality.

\begin{figure}[h!]
\centering 
\includegraphics[width=0.40\textwidth]{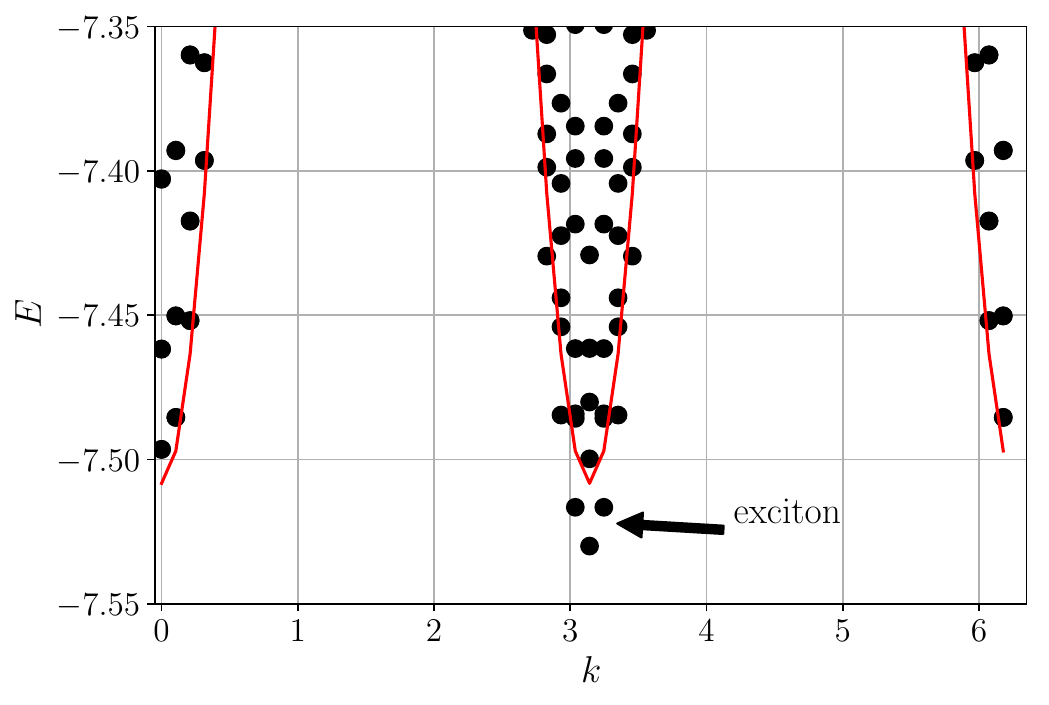}
\caption{The low-energy part of model (\ref{eq:perturb_sec_order}) for $U=10$. The red line marks the 
lower border of the scattering continuum. Around $k\sim \pi$ one state lies below this border, i.e. the exciton.}
\label{fig:scatter}
\end{figure}

Excited states are $k=\pi$, $k=0$ (anti-binding), $k=\pi\pm 2\pi/L$ (not-binding), $k=\pm 2\pi/L$ (anti-binding) and other momenta
localized near $\pi$ and $0$. 

By switching on a finite nearest-neighbour repulsion $V$ doublon and holon form a bound state immediately, the 
ground state with momentum $k=\pi$ lowers its energy according to the quadratic scaling behavior $E_B \propto -V^2, \xi \propto 1/V$
 and therefore $E_B \propto -\xi^{-2}$ as in the quasi-1D case on the triangular lattice (see Fig.~\ref{fig:scatter}). 
 
\begin{figure}[h!]
\centering
\includegraphics[width=0.35\textwidth]{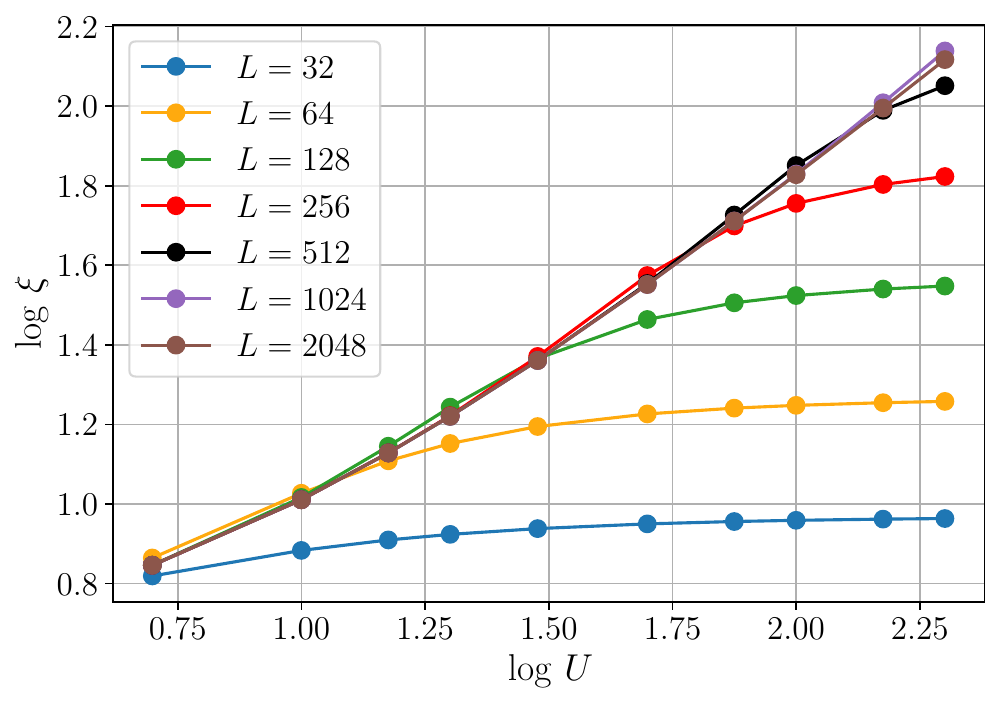}
\includegraphics[width=0.35\textwidth]{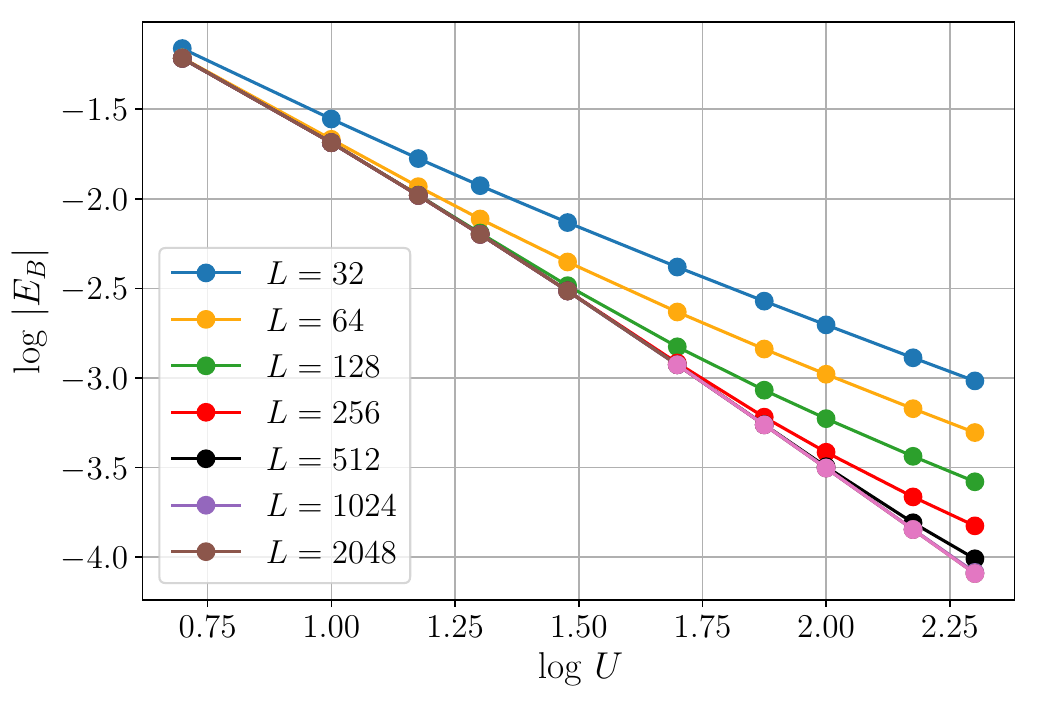}
\caption{Logarithm of the ground state binding length and binding energy as a function of the logarithmic interaction 
of model (\ref{eq:perturb_sec_order}) on the two-leg ladder with $\pi$-flux.}
\label{fig:secOrder_Ladder}
\end{figure}

In second order the effective Hamiltonian is (\ref{eq:perturb_sec_order}).
Diagonal contributions generate a static binding with amplitude $-t^2/(2U)$, illustrated in Fig.~\ref{fig:secOrder_statBindLadder}, which is the same on the triangular 
 and square lattice. Even on a $1d$ chain the amplitude doesn't change since the relative change of bonds between both configurations is independent of the geometry.
Diagonalization of (\ref{eq:perturb_sec_order}) yields the binding behavior shown in Fig.~\ref{fig:secOrder_Ladder}. The convergence behavior is much better than in the $2d$ case.
At $U=10$, system sizes of $L=128$ are sufficiently large to describe the thermodynamic limit. Even at $U=100$, $E_B \sim 10^{-4}$, the saturation is reached for $L=1024$. 
With increasing interaction the binding gets weaker with the scaling $E_B \propto -U^{-2}$ for large $U$. Hence, comparing with the quasi-$1d$ triangular lattice 
and up to second order, the Hubbard interaction acts the same as the inverse of the repulsive nearest-neighbour interaction, $U \sim 1/V$.

The qualitative change in the spectrum can be seen in Fig.~\ref{fig:scatter} (bottom). It has less overlapp with the free scattering problem (black circles) and 
bound states appear below the lower border of the scattering continuum (red line) around the $k=\pi$ dip. The bound states can be lowered down by switching on $V$.
 
In the CMI phase the ground state is doubly degenerate and consists of two staggered current configurations. 
As on the square lattice but other than on the triangular lattice the unit cell of the currents is not equal 
the unit cell of the lattice, it's reproduced by translating two lattice spacings and hence the 
lattice translation symmetry is broken. By performing a symmetry analysis one finds that the representation decomposes into $(k=0).A$ and $(k=\pi).B$, where 
$A$ ($B$) denotes symmetric (antisymmetric) under reflection along the plane perpendicular to the legs, respectively. 

Returning to the picture in Fig.~\ref{fig:exciton_state}, 
we see hints of binding and the first neutral excited state above the Mott ground state is an exciton, 
which forms the degenerate ground state together with the trivial Mott state $(k=0).A$ and breaks parity,
translation symmetry and time-reversal symmetry in the thermodynamic limit.

We conclude the first part of the paper and discuss the prospects of the Chiral Mott phases in 1 and 2D 
from our perturbative viewpoint. In second order perturbation theory terms 
are generated which create a microscopic glue between doublon and holon for purely Hubbard interactions. In two dimensions the binding is very weak and therefore within this perturbative view the CMI phase is not expected to appear in a wide parameter window without additional interactions.
A repulsive nearest-neighbour interaction $V$ creates binding but with an exponential scaling for small $V$ in $2D$. 
Further, large $V$ is problematic since the Mott insulator becomes unstable and a transition to a charge-density-wave may arise.
In $1D$ the binding is strongly enhanced and the scenario for the CMI much more realistic.

From a field theoretical point of view, the ground state of a $1+1$D Bose-Hubbard model with integer filling can be mapped to a classical $2$D XY model
where temperature takes the role of the tuning parameter \cite{Paramekanti2012}. The quantum phase transition(s) from the chiral superfluid to the conventional 
Mott insulator thus have a classical analogon in $2$D. Since the XY model is realized in Josephson-junction arrays in a magnetic field it is experimentally relevant \cite{josephson_junction_arrays}.
For the fully frustrated case Monte Carlo simulations give evidence for two very close but separate transitions 
which enclose the classical analog of the chiral Mott insulator \cite{hasenbusch_XY}.
Further, the data indicates the existence of a zero-temperature multicritical point.

In three dimensions the situation is different. Monte Carlo and real-space renormalization group calculations on
 a simple cubic lattice show only one second order phase transition \cite{kwangmoo_3dxy}.
Though, for the $3$D XY stacked triangular antiferromagnet only a weak first order behavior is found \cite{PhysRevB.69.220408}. 
Even for XY antiferromagnets on simple cubic lattice  with two additional intralayer exchanges solely one first order transition appears \cite{Sorokin2011}.
These results further underline the difficulty to realize large windows of chiral Mott insulators in 2d geometries. 

In the following part we focus on the one-dimensional case and determine the intermediate CMI regime  in different parameter and filling regimes
directly using the density matrix renormalization group (DMRG).

\section{DMRG simulations on ladder geometries}

\subsection{$\pi$-flux 2-leg square ladder}

In this subsection we perform finite-size DMRG simulations \cite{white_dmrg} on the 2-leg ladder with $\pi$-flux and
reassess the small CMI phase \cite{Paramekanti2012,Paramekanti2013} directly. 
The three phases found by Dhar \textit{et al.} (related to scenario two in Fig.~\ref{fig:phasediagram})
correspond to the vortex-lattice superfluid (VL$_{1/2}$-SF), vortex-lattice Mott insulator (VL$_{1/2}$-MI) aka CMI and 
the Meissner-Mott insulator (M-MI), mentioned in the later literature \cite{greschner_reversalCurrent,PhysRevA.94.063628}.
We use 12-15 finite-size sweeps with up to 1200 states.

\subsubsection{\label{lab:bindingenergy}Binding energy}

\begin{figure}[h!]
\centering 
\includegraphics[width=0.45\textwidth]{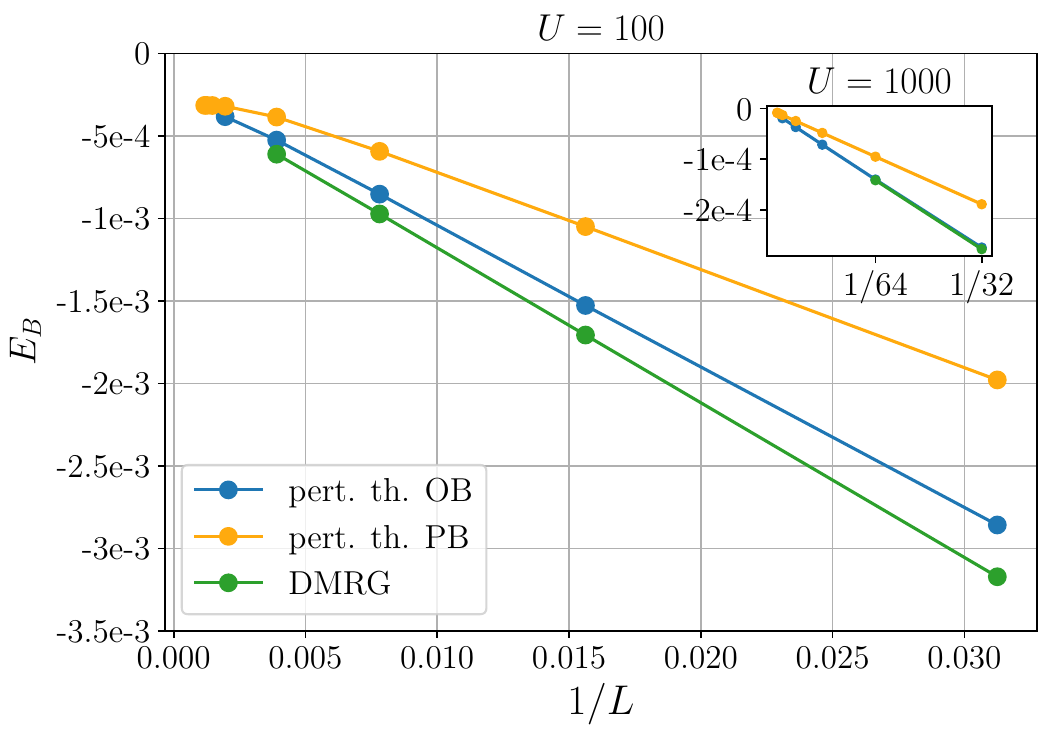}
\caption{Binding energy of the first excited state as a function of system size
 at $U=100$ and $U=1000$ (inset), calculated with DMRG and 2nd order perturbation theory.}
\label{fig:dmrg_comparison_perturb}
\end{figure}

\begin{figure}[h!]
\centering 
\includegraphics[width=0.40\textwidth]{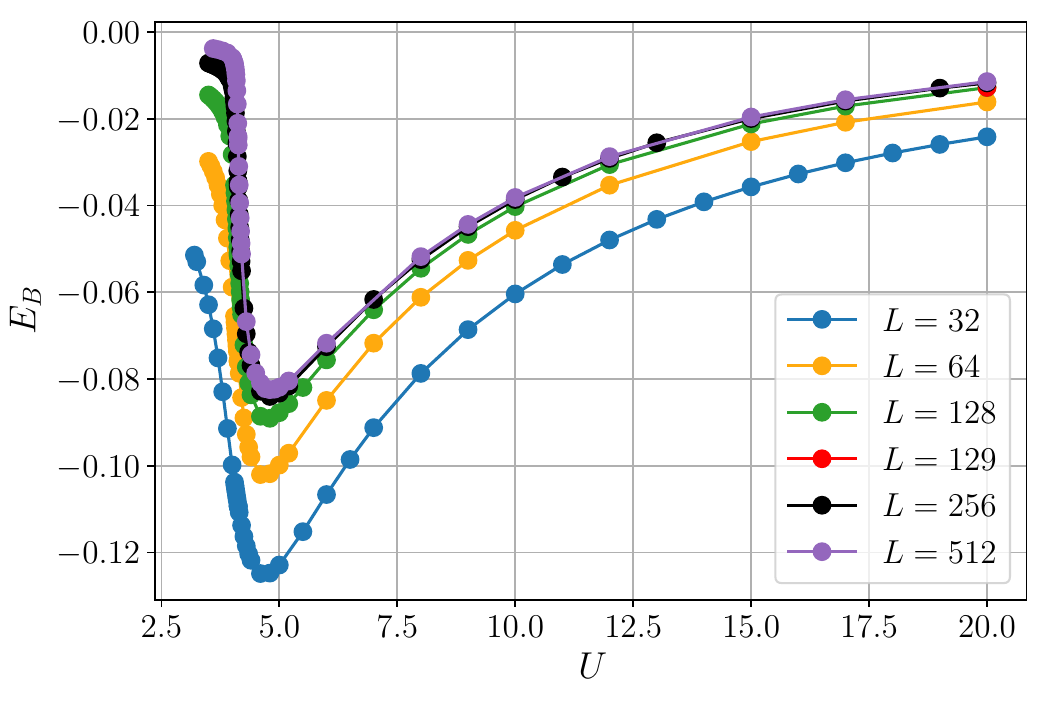}
\caption{Binding energy of the first excited state on the two-leg ladder with $\pi$-flux as a function of interaction. }
\label{fig:ladder_EB}
\end{figure}

Before we analyze the precise location of the phase transitions we test the perturbative result and compare it
with DMRG calculations. We calculate the binding energy as defined in Eq. (\ref{eq_bindingenergy}) by running three simulations for $N-1,N$ and $N+1$ particles,
 where we use two target states for the system with $N$ particles to capture the first neutral excited state.
Binding energy results of DMRG and the perturbation theory are shown in Fig.~\ref{fig:dmrg_comparison_perturb} as a function of the inverse system size. 
For large enough system sizes the difference vanishes between open boundaries and periodic boundaries calculated with perturbation theory.
Despite the large $U=100$ a surprisingly large deviation exists between second order perturbation theory and DMRG.
This missleading fact arises from the very small binding energy. A direct comparison of the first excited state for $L=32$ and $U=100$ gives 
a relative error $<0.1\%$.
The discrepancy of $E_B$ drops by going one order of magnitude higher in $U\sim 1000$ (see the inset of Fig.~\ref{fig:dmrg_comparison_perturb}). 
Nevertheless the finite-size behavior is the same in both cases if open boundaries are used.
 The binding energy as a function of $U$ is shown in Fig.~\ref{fig:ladder_EB}. On the largest system size $L=512$, $E_B$ is negative and almost saturated,
which proves the existence of an exciton as the first excited state. $|E_B|$ increases with decreasing $U$ and drops suddenly after reaching a maximum at 
$U\sim 4.8$, since near criticality the particle-hole gap changes its behavior and gets smaller more rapidly.

\subsubsection{\label{sec:BKT}BKT transition}

The superfluid-Mott transition in one-dimension is of the BKT type \cite{Kosterlitz73,0022-3719-7-6-005}.
In order to detect this transition ($U_c$) we measure the particle-hole gap $\Delta$ (see Eq. (\ref{eq_bindingenergy})).
In the Mott insulating phase the system is gapped, i.e. a finite amount of energy
is necessary to overcome the interaction $U$ and thus $\Delta$ is finite.
In the superfluid phase bosons condense in the lowest energy state and can be excited without paying any energy cost, the system is gapless.
Since on a finite system $\Delta$ is never exactly zero for $U>0$, we have to perform a finite-size scaling in order to determine the phase transition.
As a first attempt we try to extrapolate the gap to infinity. This means we expand $\Delta$ in terms of $1/L$. Truncating the series at order two yields
\begin{equation}
 \Delta(1/L) = a_0 + \frac{a_1}{L} + \frac{a_2}{L^2} + O(\frac{1}{L^3}).
\end{equation}
\begin{figure}[h!]
\centering 
\includegraphics[width=0.40\textwidth]{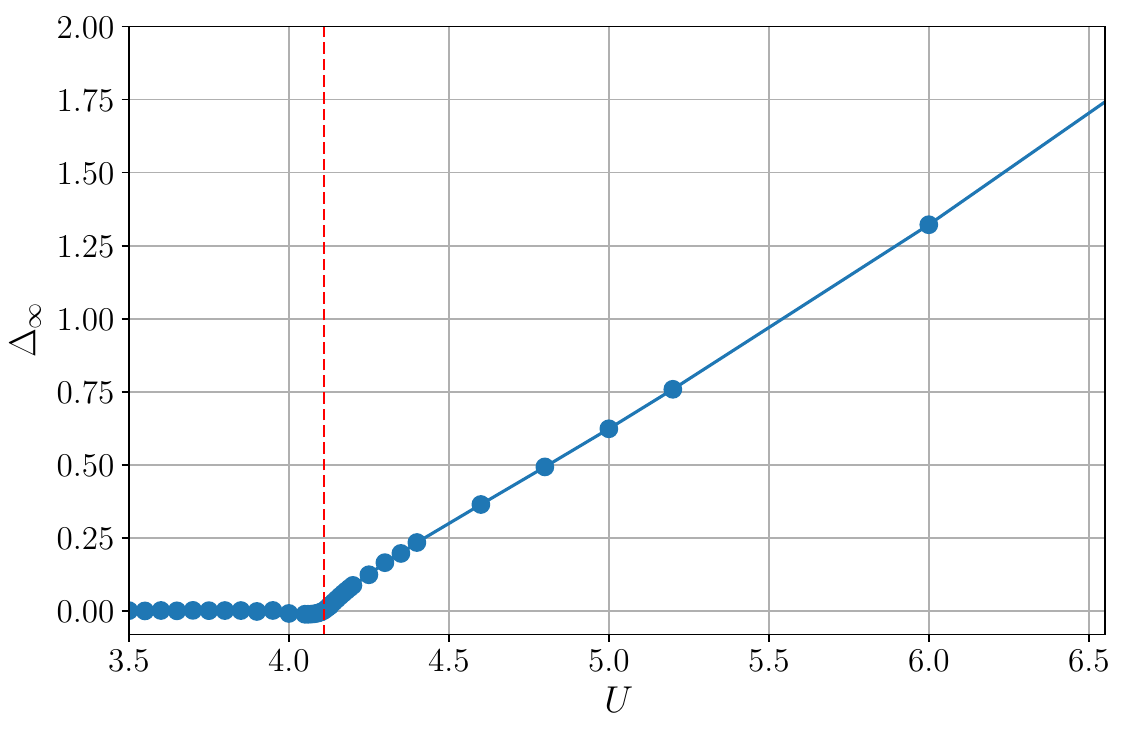}
\caption{Extrapolated particle-hole gap of the Bose-Hubbard model on the two-leg ladder with $\pi$-flux as a function of interaction. The vertical dashed red line
signalizes the crossing with the $U$-axis at $U_{c}=4.11$.}
\label{fig:ladder_gapextrapol}
\end{figure}

By fitting this function to the plot $\Delta$ vs. $1/L$ one can determine the coefficients and read off the extrapolated gap value $\Delta_\infty=\Delta(0)=a_0$.
Repeating this procedure for every $U$ yields $\Delta_\infty(U)$, shown in Fig. \ref{fig:ladder_gapextrapol}.
The red vertical line signals the $U_{c}=4.11$ value where in the thermodynamic limit the gap is barely zero, i.e. the BKT transition.
Below the transition the fit procedure, for large enough system sizes,
 gives approximately $a_1=1$, $a_0=a_2=0$, reproducing the expected scaling behavior $\Delta \sim 1/L$ for a Luttinger liquid.
Towards strong coupling $U>U_{\text{BKT}}$ the particle-hole gap shows a linear dependence $\Delta \sim U$ in agreement with the Mott regime.
Noticeable are the negative gap values right below the transition which is an artefact of the fit procedure: due to strong finite size effects $\Delta(1/L)$ seems to be 
concave on the given system sizes and yields a negative gap with a quadratic fit. In fact, larger $L$ hide the true convexness of the curve. 
Furthermore, due to the expected exponential closing of the gap \cite{0022-3719-7-6-005} $\Delta \sim e^{-\frac{b}{\sqrt{U-U_{c}}}}$ small errors on the gap 
values lead to large errors for a estimated 
transition. To get a more precise result we go beyond the simple extrapolation and use a gap scaling ansatz \cite{mishra_gapscaling,marcello_gapscaling}.

\begin{figure}[h!]
\centering 
\includegraphics[width=0.45\textwidth]{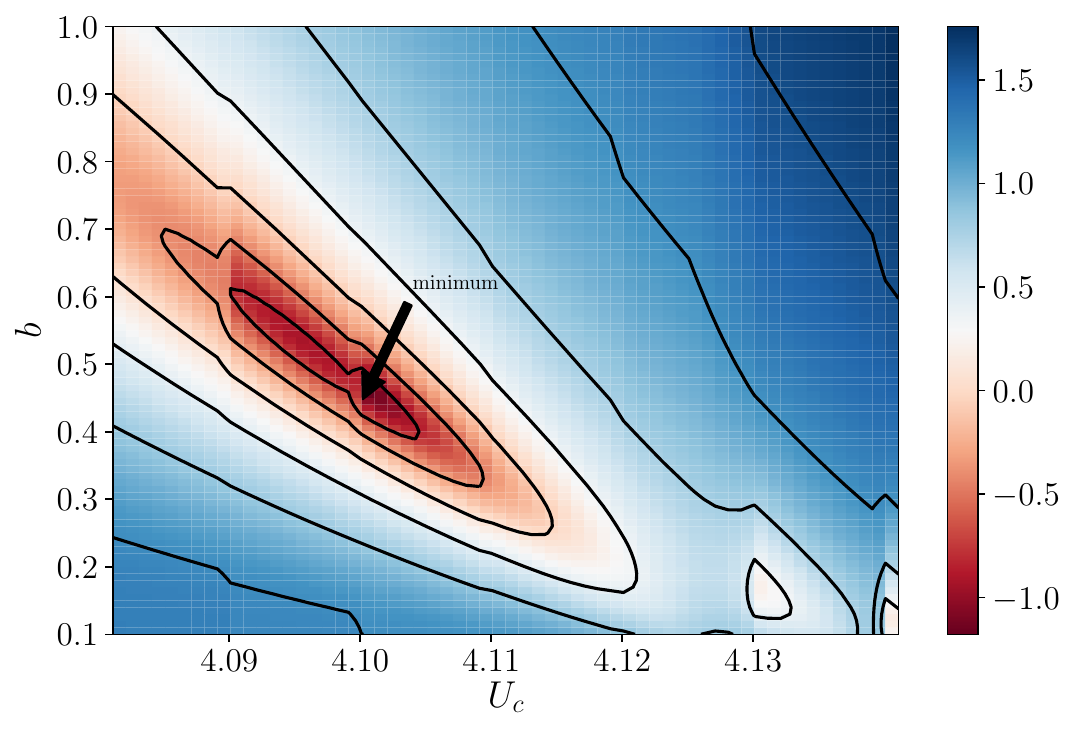}
\includegraphics[width=0.42\textwidth]{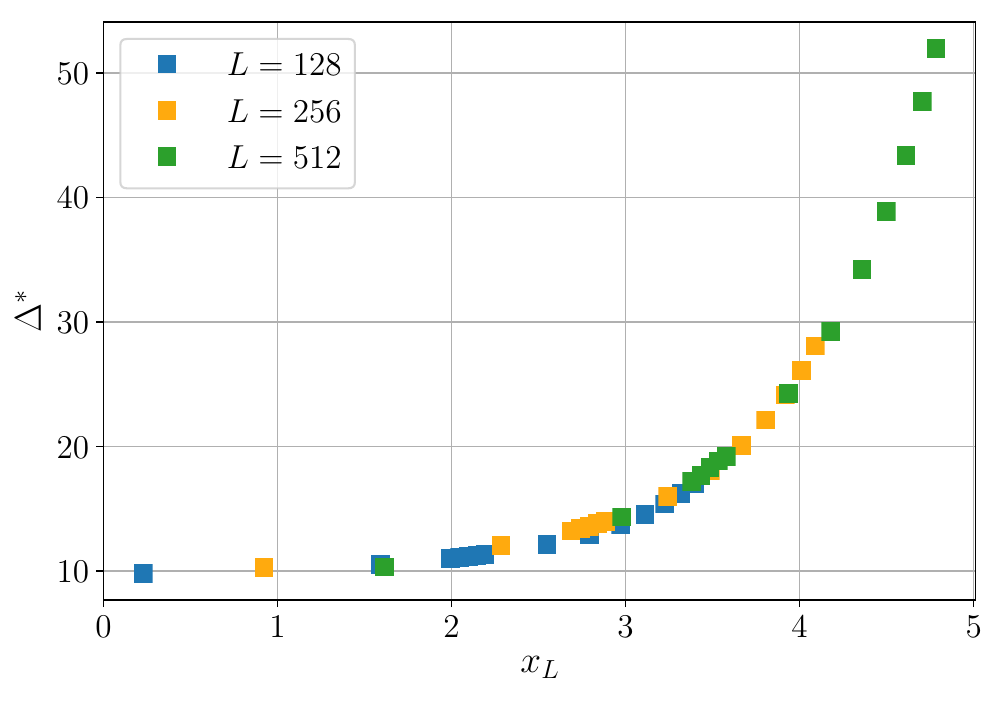}
\caption{Result of the gap scaling ansatz. Logarithm of the sum of squared residuals $\log S$ as a function of $b$ and $U_c$ at $C\rightarrow\infty$ (above). 
The minimum at $U_c=4.10$, $b=0.44$ gives
the best collapse for $\Delta^*$ as a function of $x_L$ (below) and corresponds to the critical point.}
\label{fig:ladder_scaling_S}
\end{figure}

We assume the finite-size behavior of the gap as follows
\begin{equation}
 \Delta^* = L \Delta \left(1+\frac{1}{2 \log L+C}\right) = F\left(\frac{\xi}{L}\right),
\label{eq:gap_scaling_ansatz}
\end{equation}
with a non-universal constant $C$ and the scaling function $F$.
Coming from the strong coupling site the correlation length diverges at the critical point and behaves as
\begin{equation}
 \xi \sim \frac{1}{\Delta} = e^{\frac{b}{\sqrt{U-U_{c}}}}.
\end{equation}
 Thus, close to the transition, $F(\frac{\xi}{L})$ is system size independent and $\Delta^*$ collapses for different $L$. The same applies for $\Delta^*$ vs. $\xi/L$ 
or $\Delta^*$ vs. $x_L$, where $x_L=\log L - \log \xi$. To determine the critical $U$, $b$ and $C$ we do the following steps: we define a small-meshed grid
$(U_c,b,C)$ and for every point on the grid we fit a function $f$ to $\Delta^*$ vs. $x_L$ for all $L$. The quality of the fit, identified as the sum of squared residuals
defines a discrete function $S(U_c,b,C)$. The minimum of this function determines the critical parameters $U_c$, $b$ and $C$. 
Fig. \ref{fig:ladder_scaling_S} shows the 
function $\log S$ and the corresponding collapse plot for the best fit at $U_c = 4.10$, $b=0.45$ and $C=\infty$,
 i.e. in the thermodynamic limit the BKT transition takes place at $U_{c} = 4.10(1)$.
We obtain for this transition a $U_{c}$ significantly larger than \citet{Paramekanti2012,Paramekanti2013} (they reported $U_{c}=3.98(1)$).

\subsubsection{\label{subsec:ladder_Ising}Ising transition}

\begin{figure}[h!]
\centering 
\includegraphics[width=0.45\textwidth]{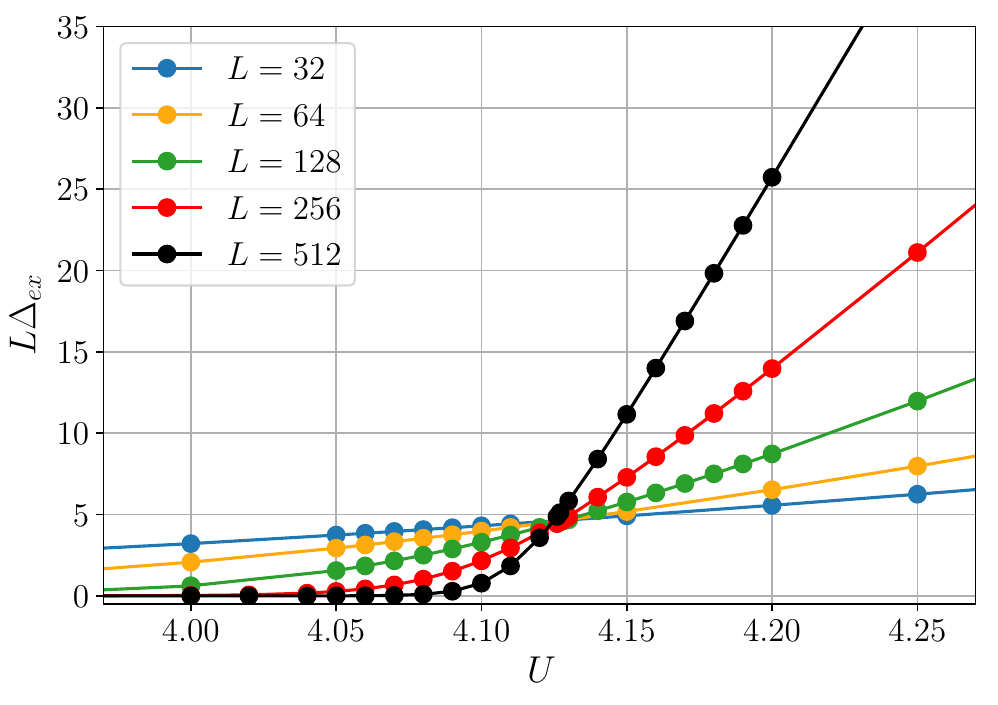}
\caption{Excitation (exciton) gap scaling of the Bose-Hubbard model on the two-leg ladder with $\pi$-flux. A clear crossing point gives evidence for a
quantum phase transition with dynamical critical exponent $z=1$.}
\label{fig:ladder_excitonL}
\end{figure}

In this section we show the results regarding the second phase transition.  As seen in Sec. \ref{lab:bindingenergy} the first excited state is really an exciton state with a finite binding energy.
We claim that this state is responsible for the Ising-like transition and developing chirality when merging with the ground state. 
We test this hypothesis by tracking the exciton gap $\Delta_{\text{ex}}$. Indeed, by plotting 
$L(E_1-E_0)$ vs. $U$ for different $L$ we discover a crossing point at $U_{cr} \approx 4.125(5)$ as shown in Fig.~\ref{fig:ladder_excitonL}.
 At this point $L(E_1-E_0)\sim \text{const}.$ in $L$ and therefore 
$E_1-E_0 \sim 1/L$ which proves a quantum critical point with the dynamical critical exponent $z=1$ \cite{sachdev2001quantum}. 
To check if it is really an Ising-like transition we 
analyze the critical exponent $\nu$. It appears in the scaling behavior of the excitation gap
\begin{equation}
 \Delta_{\text{ex}} = E_1-E_0 \sim (U-U_c)^{z\nu}.
\label{eq:ising_nu}
\end{equation}
Relation (\ref{eq:ising_nu}) is only valid in the thermodynamic limit. We extrapolate the system sizes to infinity by the same procedure as shown in Sec. \ref{sec:BKT}
and get the function $\Delta_{\text{ex},L=\infty}(U)$. By plotting $\log \Delta_{\text{ex},L=\infty}$ as a function of $\log(U-U_c)$ and fitting a line to it, we can identify 
the slope with the exponent in relation (\ref{eq:ising_nu}) and obtain $z\nu\approx 1$, which corresponds indeed to the $2D$ Ising universality class with
 $z=1$ and $\nu=1$.

\begin{figure}[h!]
\centering 
\includegraphics[width=0.48\textwidth]{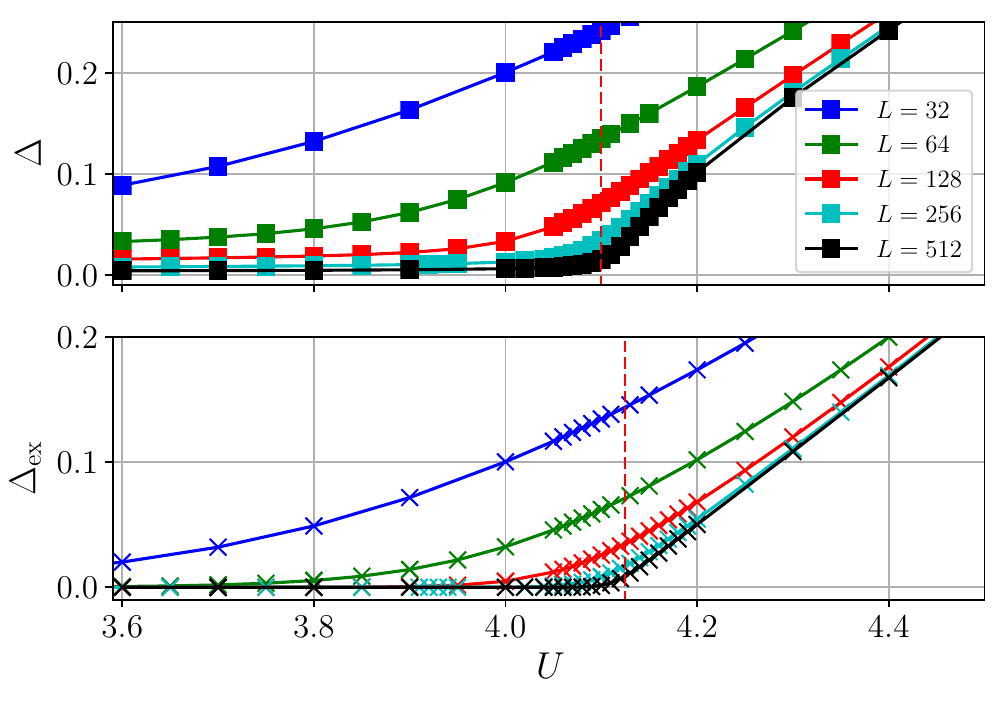}
\caption{(top) Particle-hole gap and excitation gap (bottom) as a function of interaction for different system sizes. The 
two vertical red dashed lines signalize the thermodynamic limit for the two quantities.}
\label{fig:ladder_excitonGap}
\end{figure}

In Fig.~\ref{fig:ladder_excitonGap} we show both the particle-hole gap $\Delta$ (square) and the excitation gap $\Delta_{\text{ex}}$ (cross) for different $L$. 
For system sizes $L\ge 128$ one can see a concave tendency in the particle-hole gap, which tightens with increasing $L$. This bending down is characteristic for 
the BKT transition and follows the functional form $e^{-\frac{b}{\sqrt{U-U_{c}}}}$. The exciton gap closes exponentially on finite sizes but on a shrinking $U$ scale for 
increasing $L$. For example, $\Delta_{\text{ex}}$ is almost converged for $L=512$ above $U\sim 4.3$ and resembles the linear behavior (\ref{eq:ising_nu}). 
However, near the critical BKT point the finite size effects of $\Delta$ are strong and in combination with the small non-universal parameter $b=0.45$ (see the 
gap scaling ansatz in Sec. \ref{sec:BKT}) they lead to a very small but finite CMI window of size $\delta \sim 0.02(1)$ (cf. \textit{Dhar. et. al.} in
Ref. \cite{Paramekanti2012} obtained for the Ising transition $U_c=4.08(1)$ and thus $\delta \sim 0.10(1)$).

\subsubsection{Entanglement entropy}

\begin{figure}[h!]
\centering 
\includegraphics[width=0.35\textwidth]{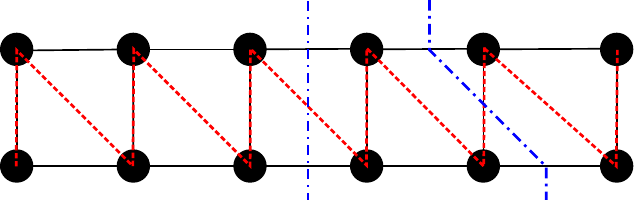}
\caption{Pictorial representation of the DMRG path (red dashed line) and the two possible cuts for a bipartition (blue dotted dashed line).}
\label{fig:ladder_picto}
\end{figure}

In this subsection we measure the entanglement entropy to get information about the CMI from another point of view.
The entanglement entropy $S$ of a system A is the Von-Neumann entropy of the reduced density matrix $\rho_A$ regarding a bipartition $A|B$ of the system.
\begin{equation}
 S_A = -\Tr(\rho_A \ln \rho_A) = -\sum\limits_{j} \lambda_i \ln \lambda_i, \; \rho_A = \Tr_B \rho,
\end{equation}
where $\lambda_i$ are the eigenvalues of the reduced density matrix which is obtained by tracing out the second partition.
Within the DMRG framework this quantity is very easy to calculate since the eigenvalues of $\rho_A$ for a certain blocklength $l$
 are calculated anyway during each renormalization step \cite{white_dmrg}. Fig.~\ref{fig:ladder_picto} illustrates the ladder with tunneling couplings in black,
the DMRG path to build up the ladder in red and the two possible types of cuts in blue. We are using the left type of cut generating two rectangular subladders.
The blocklength $l$ is then defined as the linear length of the left subladder. The length of the right subladder is respectively $L-l$.
 For example, the partitioning in Fig.~\ref{fig:ladder_picto} has a blocklength of $l=3$.

\begin{figure}[h!]
\centering 
\includegraphics[width=0.48\textwidth]{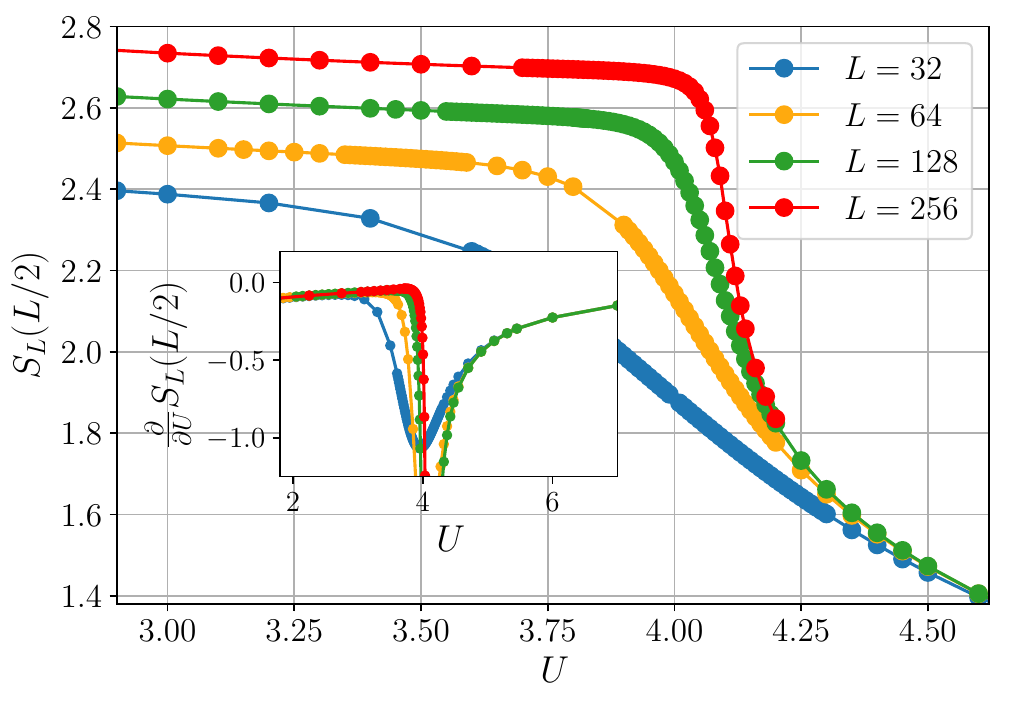}
\caption{ Entanglement entropy of the Bose-Hubbard model on the two-leg ladder with $\pi$-flux as a function of interaction. (inset) Derivative of the entropy.}
\label{fig:ladder_entropy}
\end{figure}

The entanglement entropy with a blocklength of $L/2$ and its derivative (inset) is shown in Fig.~\ref{fig:ladder_entropy}.
 A sudden drop around $U\sim 4$ signalizes a drastic
qualitative change of the wavefunction. Right to the drop, the correlation length is finite implying a saturation of the entanglement since particles
separated from each other by a distance much larger than the correlation length are almost uncorrelated and therefore no information is exchanged.
The saturation can clearly be seen and is a proof for the Mott insulator phase. Below the drop, doubling the system size leads to a constant increase of
entanglement, which is a strong hint for superfluidity in one dimension (see Eq. \ref{eq:log_entropy}). 
With increasing $L$ the breaking down of the entropy shifts towards larger $U$,
visible in the derivative as a negative peak. Further, close to the drop, the derivative increases and approaches nearly zero. 
It is not clear if for even larger system sizes the derivative exceeds zero and develops a positive peak.

Another characterization for critical points and the corresponding phases (universality classes) is via the central charge.
In the vicinity of a quantum critical point, the correlation length is much larger than the lattice spacing $\xi \gg a$ and the system is called \textit{critical}. 
The low energy physics of a one-dimensional system is then described in the continuum limit by a quantum field theory in $1+1$ dimensions.
At the critical point the system is conformal invariant and described by a conformal field theory with central charge $c$ \cite{HOLZHEY1994443,cardy_cft}.
It has been shown that the entanglement entropy for a critical system with open boundary conditions is given by \cite{cardy_cft}
\begin{equation}
  S_L(l) = \frac{c}{6} \log \left[ \frac{2L}{\pi} \sin \left(\frac{\pi l}{L}\right)  \right] + \log g + c_1/2,
\label{eq:log_entropy}
\end{equation}
where g is the boundary entropy \cite{affleck_cft} and $c_1$ a non-universal constant. For the Bose-Hubbard model, the system at the
BKT transition and even in the whole gappless phase is described by a Tomanaga-Luttinger
 liquid \cite{Giamarchi:743140} with $c=1$. For the second phase transition, which is expected in the Ising universality class, the central charge amounts $c=1/2$.
 In the gapped Mott insulator phase the central charge is zero.

\begin{figure}[h!]
\centering 
\includegraphics[width=0.40\textwidth]{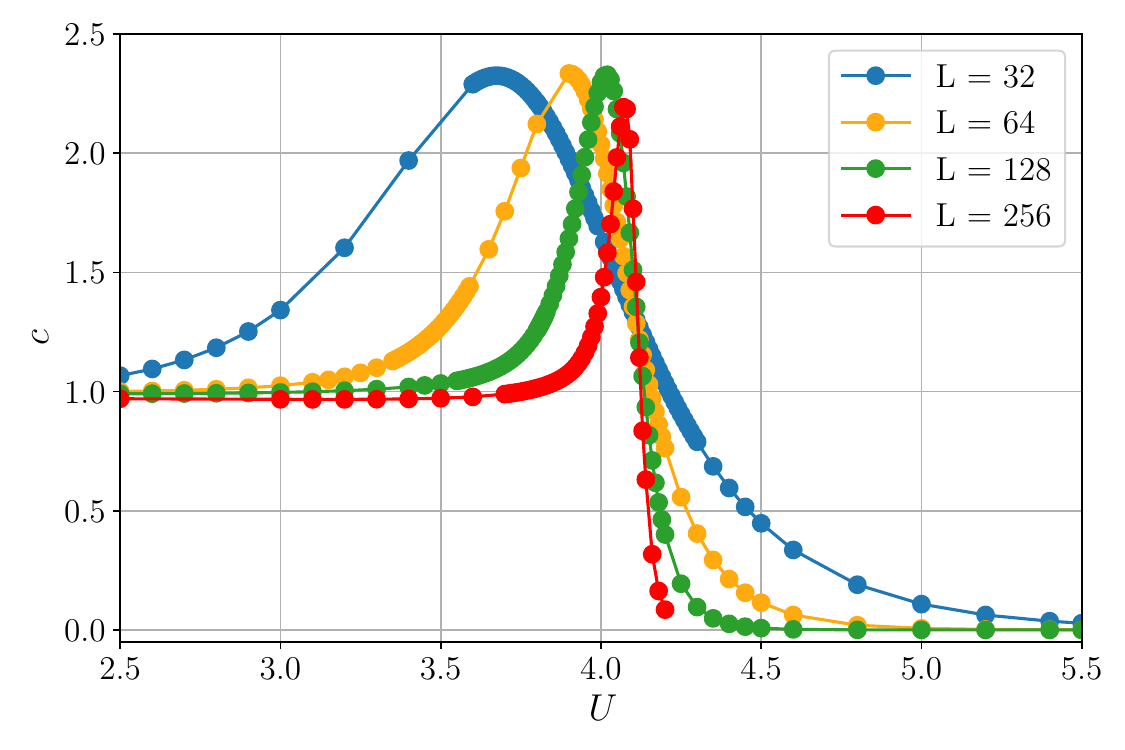}
\caption{Central charge of the Bose-Hubbard model on the two-leg ladder with $\pi$-flux as a function of interaction, extracted from the entanglement entropy.}
\label{fig:centralcharge}
\end{figure}

One way to measure the central charge is a direct fit of the function (\ref{eq:log_entropy}) to the data $S_L$ vs. $l$. The coefficients $g$ and $c_1$ are not
important and can be combined to an offset. Fig.~\ref{fig:centralcharge} shows $c$ as a function of $U$. For small interactions $U$ the central charge approaches $c=1$ 
in agreement with the analytical prediction for a Luttinger liquid. Towards strong coupling $c$ drops off to zero very quickly. Strikingly is the peak behavior in
the critical region with values of $c>2$. Merging of both phase transitions would give $c=3/2$, which is quite smaller than the values of $c$ in the data.
However, $c$ starts to reduce its peak value for $L \ge 256$. We conclude that the system sizes are too small to resolve the small CMI window.

\subsection{Other geometries at unit filling}

\begin{figure}[h!]
\centering 
\includegraphics[width=0.40\textwidth]{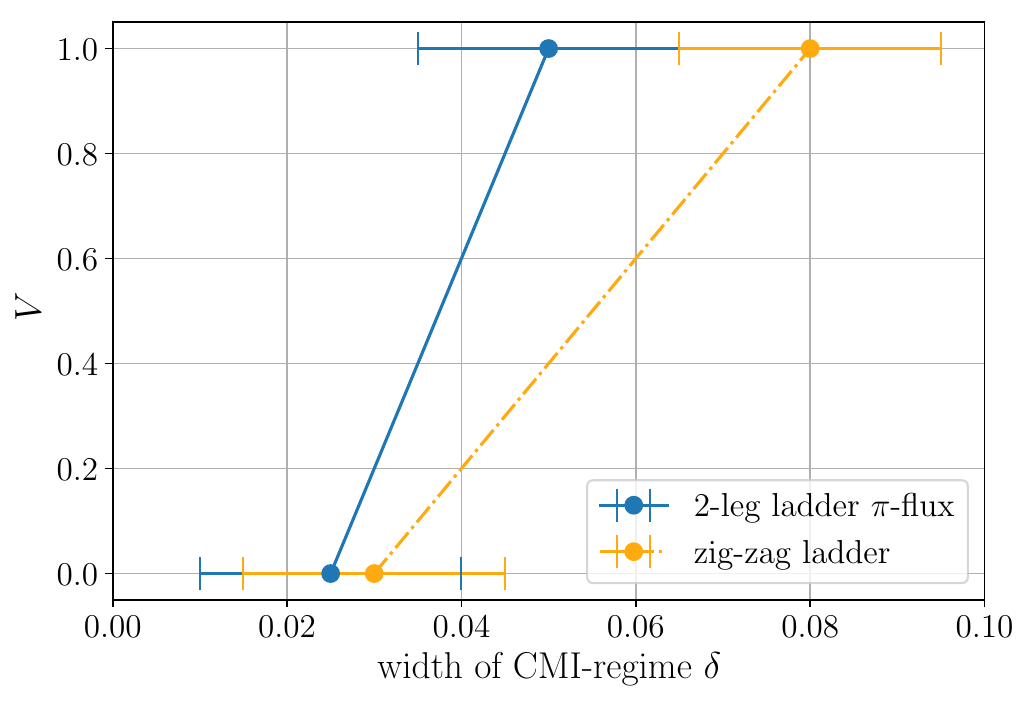}
\caption{Schematic representation of the enlargement of the CMI-regime with repulsive nearest-neighbour interaction $V$ for the 2-leg ladder with $\pi$-flux and zig-zag ladder.}
\label{fig:widthCMIregime}
\end{figure}

Before we study the anisotropic case at half-filling we show our results of the same analysis on the zig-zag ladder in Tab. \ref{tab1} and Fig.~\ref{fig:widthCMIregime}. 
The chiral Mott phase has almost the same extension on the zig-zag ladder, very narrow but finite. Switching on a repulsive nearest-neighbour interaction $V$ directly
lowers the energy of the exciton \cite{zalatel2014}. Indeed this lowering leads to an enlargement of the CMI although on small scale: an interaction of $V=1$ enlarges the phase on the square ladder
to a width of $\delta \sim 0.050(15)$, which is still small. On the zig-zag ladder the enlargement is enhanced giving $\delta \sim 0.08$.
 This qualitative enlargement is consistent with Ref. \cite{zalatel2014}.
Another way to manipulate the extension of the phase would be to change the hopping ratio $t_\perp/t$. Reducing the ratio on the square ladder can lead to a tricritical point 
with direct CSF-MI transition (scenario one in Fig.~\ref{fig:phasediagram}), which was shown by a field-theoretical approach \cite{artificialMagFieldFieldTheoretical}.
Alternatively, changing the density to $\rho=0.5$ and increasing the ratio $t_\perp/t$ leads to a huge
 CMI (vortex-lattice $\mathrm{VL}_{1/2}$ with vortex density $\rho_V=1/2$) region as Greschner \textit{et al.} showed \cite{greschner_reversalCurrent,PhysRevA.94.063628}.

\begin{table}[h]
\centering
\begin{tabular}{c|c|c|c|c}
 lattice & $V$& $U_{\text{Ising}}$ & $U_{\text{BKT}}$ & $\delta=U_{\text{Ising}}-U_{\text{BKT}}$ \\
\hline\hline
 2-leg ladder $\pi$-flux & $0$  & $4.10(1)$ & $4.125(5)$  & $0.025(15)$ \\
 2-leg ladder $\pi$-flux & $1$ & $4.22(1)$  & $4.270(5)$ & $0.050(15)$ \\  
\hline
zig-zag ladder  & $0$  & $4.19(1)$ & $4.220(5) $  & $0.030(15)$ \\
 zig-zag ladder & $1$  & $4.50(1)$ & $4.580(5) $  & $0.080(15)$ \\  
\end{tabular}
\caption{\label{tab1}Enlargement of the chiral Mott insulator phase with repulsive nearest-neighbour interaction $V$
 on the 2-leg square ladder with $\pi$-flux and the zig-zag ladder.}
\end{table}

\subsection{Rung Mott insulator at $t_\perp/t=2$}

To show really a textbook example of a broad CMI phase with substantial separation of the two phase transitions we perform DMRG simulations for 
the 2-leg ladder with $\pi$-flux, half-density and anisotropic couplings $t_\perp/t=2$.
In the hardcore boson limit $U\rightarrow\infty$, at half-filling, if the interchain coupling is very large, $t_\perp \gg t$, bosons are pinned to the rungs of the ladder,
one boson for each rung, and form a product of singlet pairs on the rungs. Adding or removing particles is always connected with a finite energy difference, i.e.
the present phase is gapped and called a \textit{rung-Mott insulator}. It turns out that even for arbitrarily small ladder coupling $t_\perp$ the system remains 
gapped in the hard-core limit\cite{crepin_rungMottHardcoreBosons}. 
Within this limit no vortex-lattice can be found and only a transition from a Meissner Mott insulator (M-MI) to vortex-liquid Mott insulator (V-MI) is possible,
where the V-MI phase vanishes above some critial ratio $(t_\perp/t)_{\mathrm{crit}}\gtrsim 1.7$ \cite{piraud_hardcoreBosons}.
For soft-core bosons the situation is different and superfluid phases as well as vortex-lattices are possible. 

\subsubsection{BKT transition}

\begin{figure}[h!]
\centering 
\includegraphics[width=0.35\textwidth]{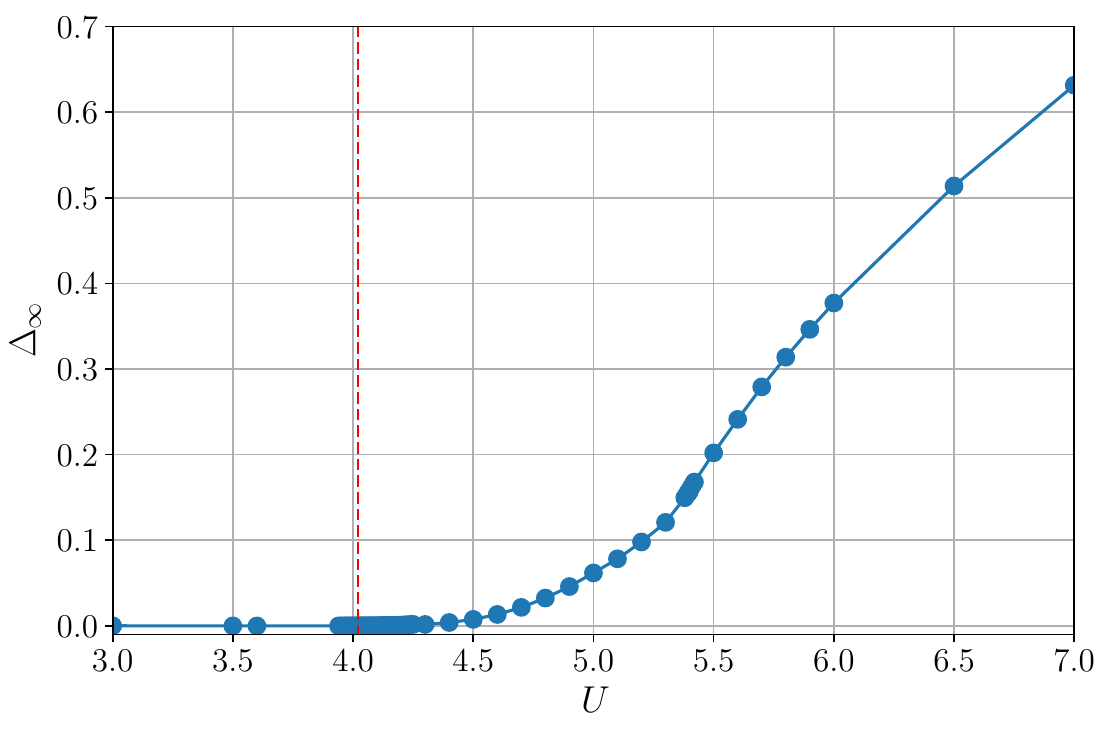}
\caption{Extrapolated particle-hole gap at half-filling and $t_\perp/t=2$ as a function of interaction. The vertical dashed red line
signalizes the crossing with the $U$-axis at $U_{\text{BKT}}=4.02$.}
\label{fig:halfdens_gapextrapol}
\end{figure}

First we identify the BKT transition to the gapless Luttinger liquid by calculating the particle-hole gap. Analogous to Sec. \ref{sec:BKT} we 
expand $\Delta$ in terms of $1/L$ and extrapolate the gap for every $U$. Fig. \ref{fig:halfdens_gapextrapol} shows the result. 
Compared to the case with density one, the gap closes much smoother and the characteristic BKT curve-form develops earlier.
 The exponential ``tail'' lasts up to $U\sim 5.4$ indicating a large value of $b$ (cf. $b=0.45$ at unity-filling) and makes the 
accurate determination of the transition more challenging. The red vertical line $U_c=4.02$ in the figure marks the last positive data point from the right.
The condition $\Delta < \epsilon = 10^{-4}$ gives $U_c = 4.06$. In this case the uncertainty is large and we can estimate the transition with $U_{\text{BKT}}=4.06(10)$.
To get a more precise answer we can again use the scaling ansatz from Eq. \ref{eq:gap_scaling_ansatz}. Accordingly, the optimal parameters are
$U_{\text{BKT}}=4.02(2)$, $b=3.8(1)$, $C=\infty$ and the sum of squared residuals per datapoint $S_{\text{min}} \sim 10^{-6}$.

\subsubsection{Ising transition}

\begin{figure}[h!]
\centering 
\includegraphics[width=0.40\textwidth]{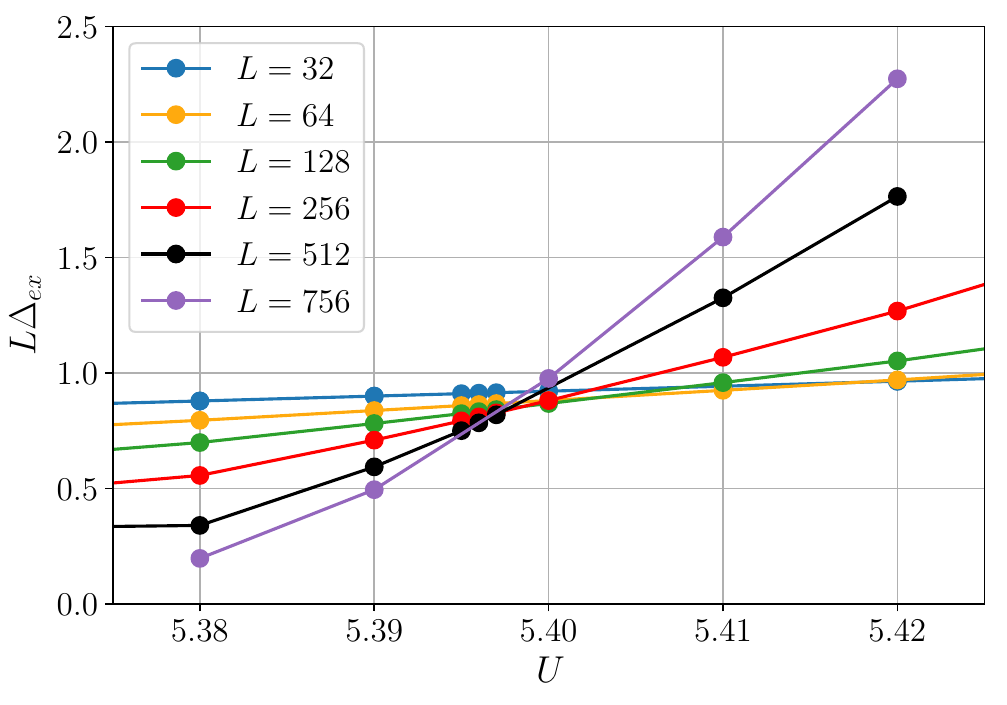}
\includegraphics[width=0.45\textwidth]{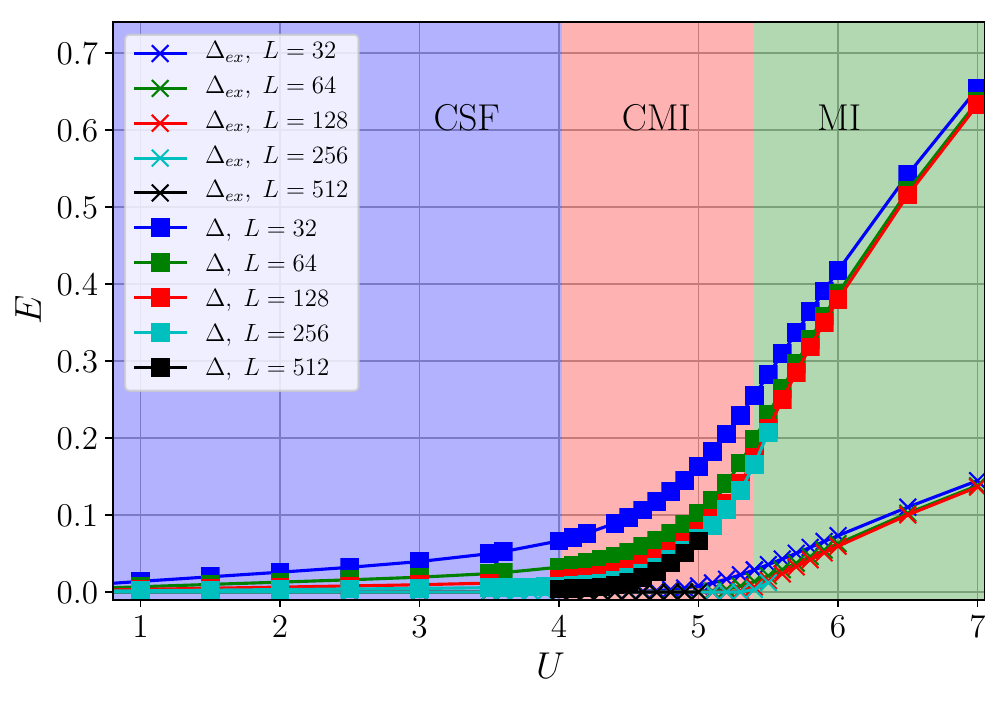}
\caption{(top) Excitation gap scaling at half-filling and $t_{\perp}/t=2$. (bottom) Particle-hole gap and excitation gap as a function of interaction
for different system sizes. The shaded red area encloses the CMI region.}
\label{fig:halfdens_ising}
\end{figure}

Turning to the second phase transition we are tracking the excitation gap (cf. Sec. \ref{subsec:ladder_Ising}). Fig. \ref{fig:halfdens_ising} (a) shows
$L \Delta_{\text{ex}}$ as a function of $U$ with a crossing point at $U_{c} = 5.40(1)$ indicating a quantum phase transition with $z=1$.
Fig. \ref{fig:halfdens_ising} (b) shows excitation gap and particle-hole gap at once for different system sizes. 
The shaded region in red denotes the CMI phase and is enclosed by points where in the 
thermodynamic limit both gaps are zero. The phase with an extension $\delta \sim 1.38(3)$ is surprisingly large compared to the density $\rho=1$ case.

\begin{figure}[h!]
\centering
\includegraphics[width=0.40\textwidth]{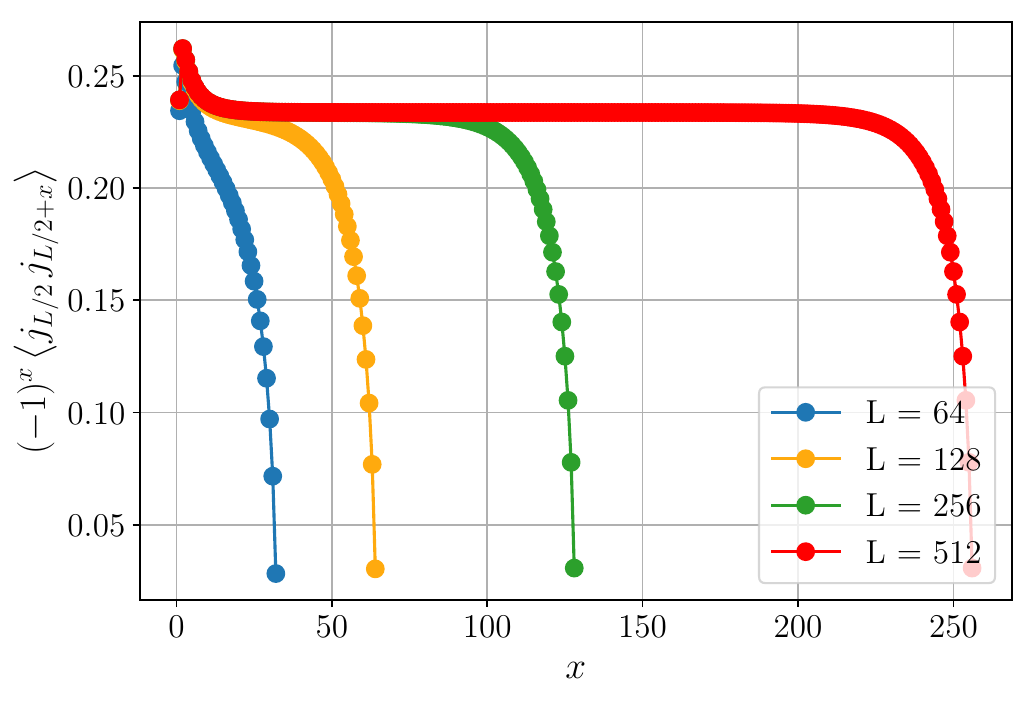}
\includegraphics[width=0.40\textwidth]{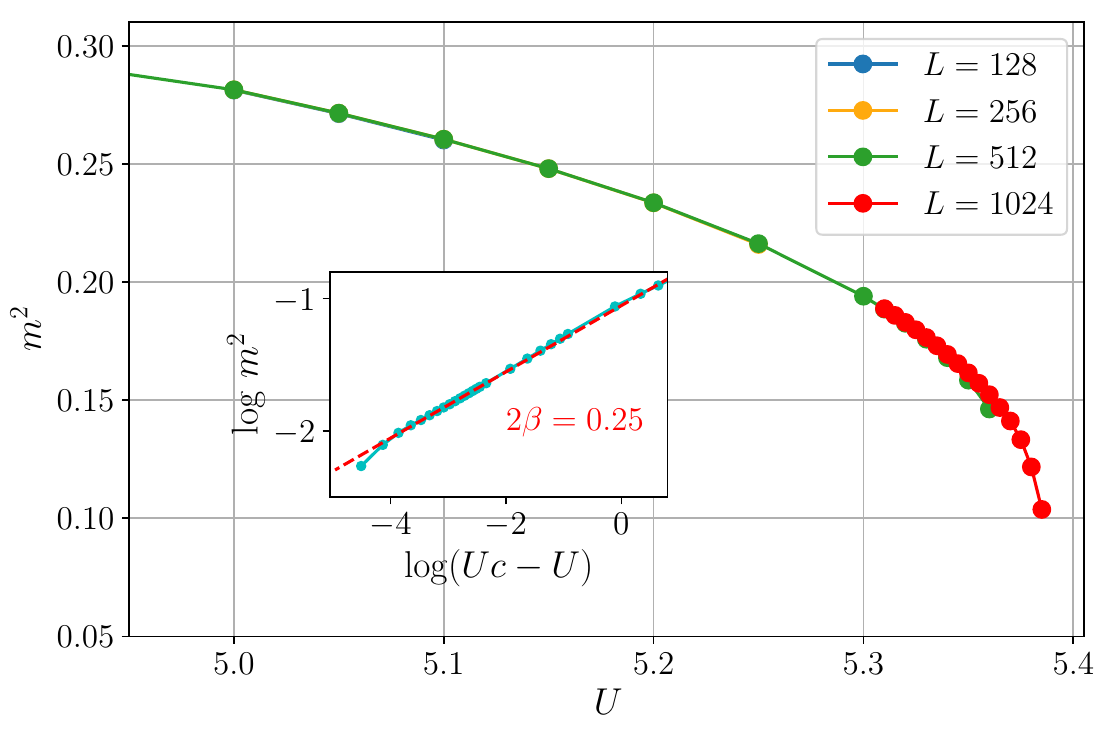}
\caption{(top) Staggered rung-rung current correlations as a function of the distance $x-L/2$ in real space. Saturation with increasing system size indicates 
long-range staggered current order. (bottom) Saturated staggered current as a function of interaction. The inset shows $\log m^2$ as a function of $\log U_c-U$,
which is a straight line with slope $2\beta=0.25$, in agreement with the Ising universality class.}
\label{fig:halfdens_isingScalingBeta}
\end{figure}

Another method to detect the phase transition is by measuring the order parameter and make a scaling-analysis.
Vortex lattice phases have finite local rung currents with a definite integer vortex periodicity \cite{piraud_hardcoreBosons}. 
The CMI is a vortex-lattice with a staggered rung-current pattern ($\rho_V=1/2$) and 
hence, one can define the Fourier transform of the rung-rung current
\begin{equation}
  S(k) = \frac{1}{L^2} \sum\limits_{l,m} e^{ik(l-m)} \left< j_l \, j_m \right>,
\end{equation}
with the current operator on the $l$-th rung $j_l=i(a_l^\dagger b_l - b_l^\dagger a_l)$,
where $a^\dagger_l$ ($b^\dagger_l$) creates a particle on the upper (lower) leg on rung $l$. 
The order parameter can be defined \cite{Paramekanti2012} as $m^2 \equiv S(\pi)$.
It is known from scaling theory \cite{sachdev2001quantum} that in the vicinity of a second order phase 
transition the order parameter scales as
\begin{equation}
 m \propto (T-T_c)^{\beta}
\end{equation}
when the transition is approached from the disordered phase $T>T_c$. In our context the interaction $U$ 
tunes the transition and corresponds to temperature in a classical model.
To overcome finite-size effects we don't take the full Fourier transform of $\left< j_l \, j_m \right>$. Instead,
by fixing $l=L/2$ and plotting $ (-1)^x \, \left< j_{L/2} \, j_{L/2+x} \right>$ as a function of $x$ one can extract 
the saturated value of $L^2\,S(\pi)$, i.e. the thermodynamic limit, as shown in Fig. \ref{fig:halfdens_isingScalingBeta} (a) for $U=5.2$.
The relaxation range for small $x$ and the decaying ``tail'' at the boundary enclose the saturated plateau sector.
The more critical the system gets the smaller becomes the plateau window.
Formally we can determine the value by defining the condition
\begin{equation}
  \partial_x (-1)^x \, \left< j_{L/2} \, j_{L/2+x} \right> < \epsilon = 10^{-4}.
\label{eq:ising_saturCond}
\end{equation}
Repeating the procedure for every $U>U_c$ yields the order parameter curve in Fig. \ref{fig:halfdens_isingScalingBeta} (b). 
Due to the increasing criticality data points very close to the transition, corresponding to small system sizes, start to deviate from
the saturation line and vanish if (\ref{eq:ising_saturCond}) can not be satisfied anymore.
We extract the critical exponent $\beta$ by plotting $\log m^2$ as a function of $\log (U-U_c)$ and fitting a line to it as shown
 in Fig. \ref{fig:halfdens_isingScalingBeta} (c).
For $U_c$ we use the estimation from the excitation gap scaling, $U_c=5.396$.
 The slope of the fit gives $2\beta \approx 0.25$ in agreement with $\beta=1/8$ of the
Ising universality class.

\begin{figure}[h!]
\centering 
\includegraphics[width=0.48\textwidth]{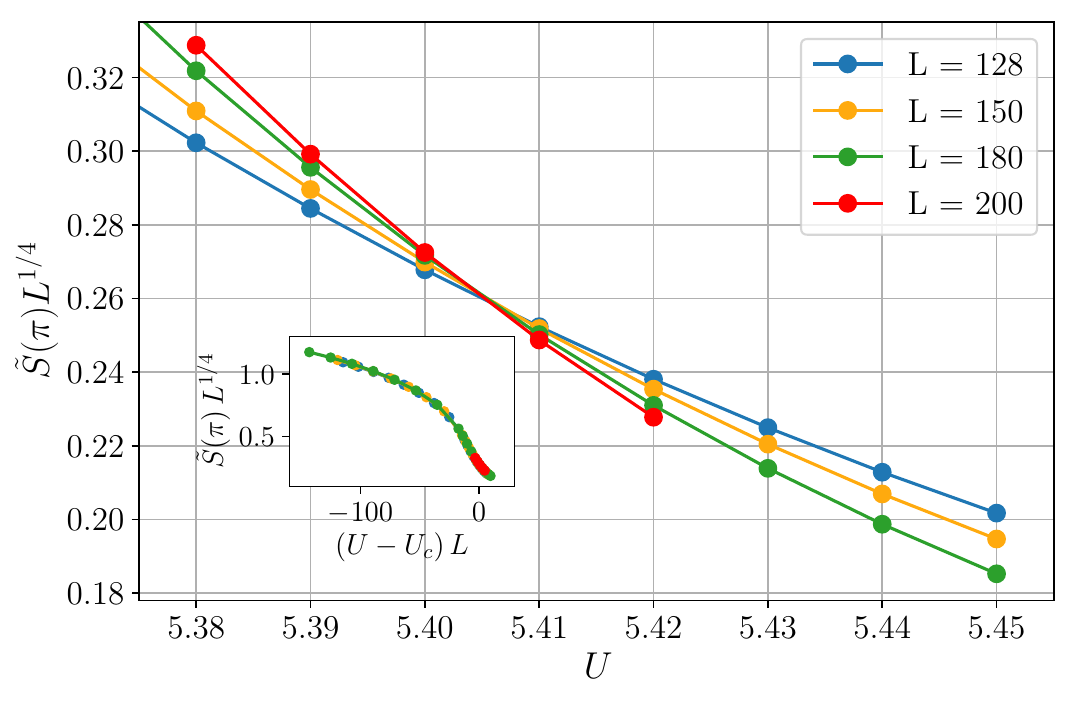}
\caption{Staggered rung-rung current scaling plot on the two-leg ladder at half-filling and $t_{\perp}/t=2$:
 $\tilde{S} (\pi) L^{1/4}$ as a function of interaction, where $\tilde{S} (\pi)$ is the staggered rung-rung current without the autocorrelation, 
$\tilde{S} (\pi) = S(\pi) - \sum_x \left< j^2_x \right> /L^2$.
 The crossing point between $U=5.40$ and $U=5.41$ is slightly drifting to smaller $U$.
 The inset shows the collapse plot, $\tilde{S}(\pi)L^{1/4}$ as a function of $(U-U_c)L$,
 for $U_c \sim 5.396$.}
\label{fig:halfdens_isingScaling}
\end{figure}

We continue by locating the phase transition with the finite size scaling of the current-current correlations.
Generally, on a second order phase transition the two-point correlation function between two sites decays algebraically \cite{sachdev2001quantum} with the critical exponent
$\eta$ 
\begin{equation}
 \left< \sigma_x \sigma_y \right> \propto \frac{1}{|x-y|^{2-d-\eta}} \propto L^{-\eta}.
\end{equation}
Further $ \frac{1}{L^2} \sum_{x,y} \left< \sigma_x \sigma_y \right> \propto L^{-\eta}$ and
 $  L^{\eta-2} \sum_{x,y} \left< \sigma_x \sigma_y \right> \propto \text{const}$, which means 
at the critical point the quantity collapses for different system sizes. 
The scaling plot $\tilde{S}(\pi)L^{1/4}$ as a function of $U$ is shown in Fig. \ref{fig:halfdens_isingScaling}, where
the autocorrelation is subtracted, $\tilde{S} (\pi) = S(\pi) - \sum_x \left< j^2_x \right> /L^2$. The crossing point 
lies between $U=5.40$ and $5.41$, slightly above the prediction from the excitation gap. Although the system sizes are smaller and the crossing
is marginally drifting towards smaller $U$. By using the finite-size scaling ansatz $\tilde{S}(\pi)L^{2\beta/\nu}=f((U-U_c)L^{1/\nu})$, where $f$ is a scaling function, we can
plot $\tilde{S}(\pi)L^{2\beta/\nu}$ as a function of $(U-U_c)L^{1/\nu}$ and see a good collapse for $U_c\sim 5.396$, shown in the inset of Fig. \ref{fig:halfdens_isingScaling}.
This confirms the Ising character of the transition with $\eta=1/4$, $\beta=1/8$ and $\nu=1$.

\subsubsection{Entanglement entropy}

\begin{figure}[h!]
\centering
\includegraphics[width=0.40\textwidth]{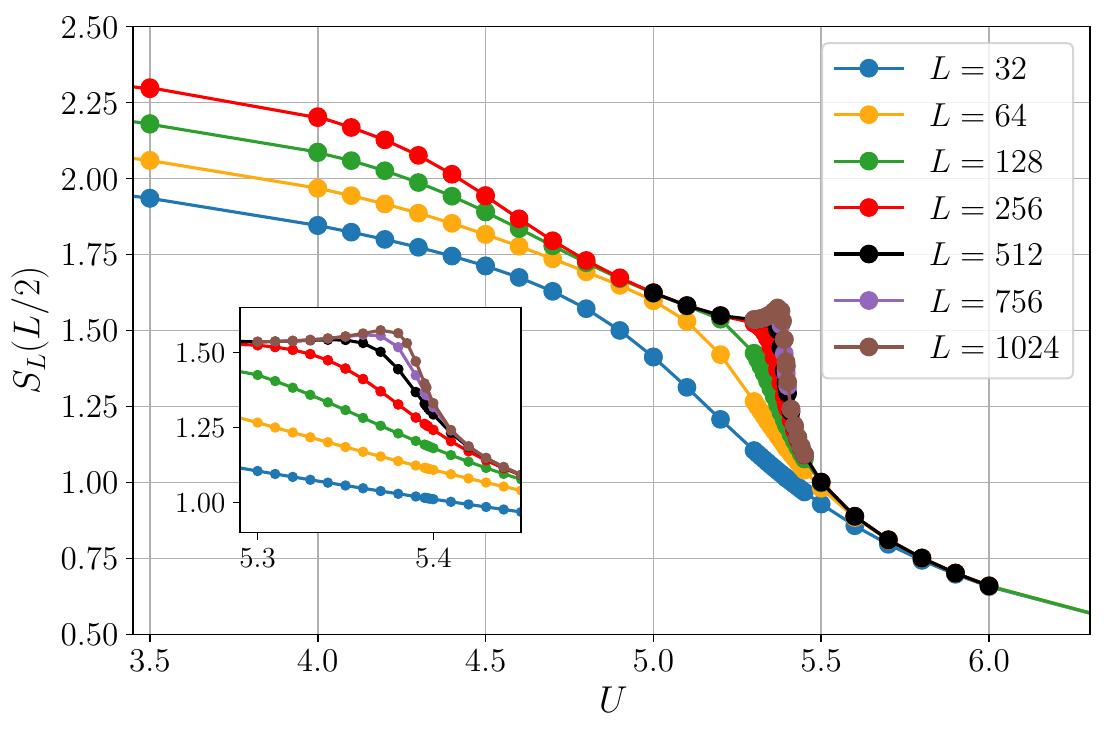}
\caption{Entanglement entropy at half-filling and $t_{\perp}/t$ as a function of interaction. (inset) Zoom around the Ising transition.}
\label{fig:halfdens_entropy}
\end{figure}

Finally we take a look at the entanglement entropy to support our estimation for the transitions. Cutting the ladder at $l=L/2$ yields
the entropy as a function of $U$ as in Fig. \ref{fig:halfdens_entropy}. Starting at $U\sim 4$ the system gets massive, with a finite correlation length, up to
$U\sim 5.35$ where system sizes split up again and a peaked feature starts to develop, evidently signal for the Ising phase transition. The inset 
zooms into the region around the transition. It's remakable that system sizes $L\ge 512$ are necessary to resolve the peak. 

\begin{figure}[h!]
\centering
\includegraphics[width=0.47\textwidth]{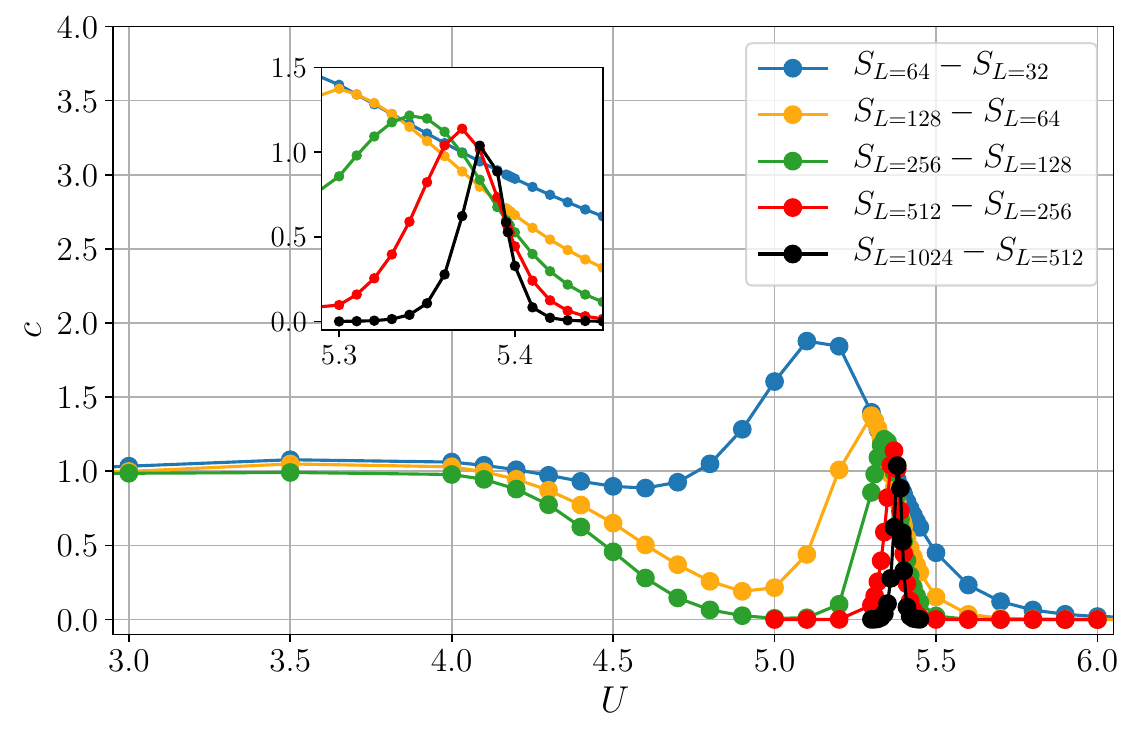}
\includegraphics[width=0.47\textwidth]{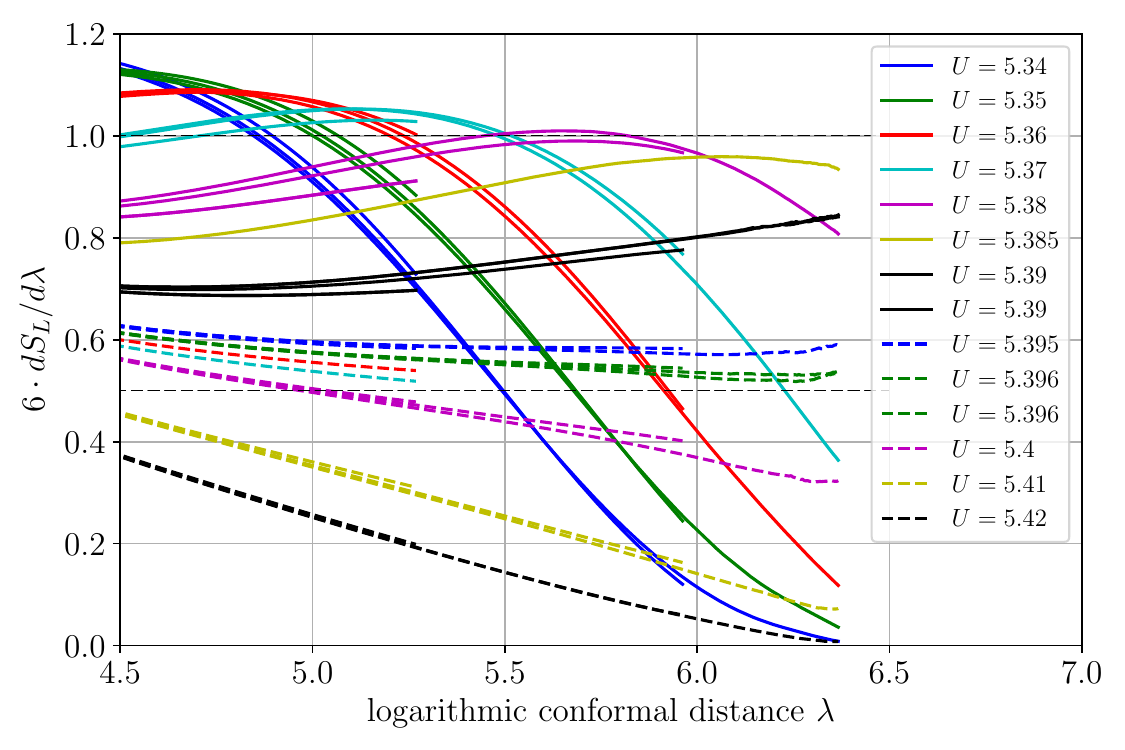}
\caption{(top) Central charge as a function of interaction $U$ determined by subtracting the entropy values of doubled system sizes (see Eq. (\ref{eq:central_charge})). 
The inset shows the vicinity of the Ising transition. (bottom) Derivative of $6 \, S_L(l)$ as a function 
of the logarithmic conformal distance $\log \lambda = \log [2L/\pi \sin (\pi l/L)]$ for different $L$ and $U$. For $U\sim 5.396$ the quantity approaches $6 \, S_L(l)=1/2$ 
for very large conformal distances.}
\label{fig:halfdens_centralcharge}
\end{figure}

Finally we extract the central charge with the following method. 
According to relation (\ref{eq:log_entropy}) the entropy maximum is $S_L(L/2)=c/6 \log(2L/\pi)+\text{const}.$ Taking 
the difference to the doubled system size gives
\begin{equation}
 S_{2L}(L)-S_L(L/2) = \frac{c}{6} \log 2 \Rightarrow c= \frac{6}{\log 2} (S_{2L}-S_L).
\label{eq:central_charge}
\end{equation}
Fig. \ref{fig:halfdens_centralcharge} (a) illustrates the result for the central charge as a function of $U$. Saturation of the entropy translates into 
$c\rightarrow 0$ and shows explicitely the Mott insulating region $U \gtrsim 4$. The developing $\delta$-function at $U\sim 5.38$ is in agreement with
the results from the energy and correlation function scaling in the previous subsection, though the peak height is not $1/2$ corresponding to the central
charge of a free fermion as predicted from CFT.
This is not a contradiction since the peak amplitude is still decreasing while the peak is slightly moving to the right.
 Fig. \ref{fig:halfdens_centralcharge} (b) shows the derivative of $6 \, S_L(l)$ as a function 
of the logarithmic conformal distance $\log \lambda = \log [2L/\pi \sin (\pi l/L)]$ \cite{andreas_entanglementSpreading} for different $L$ and $U$.
Three different function types are visible: first, for $U\lesssim 5.39$ the entropy exhibits a ``bubble''-shape by decreasing-increasing-decreasing, where 
$6 \partial_\lambda S_L \rightarrow 0$ for $\lambda \gg 1$. For larger $U$ the ``bubbles'' move to larger $U$.
Second, for $U \gtrsim 5.40$ the function directly decreases monotonously to zero.
Third, around $U\sim 5.396$ the derivative decreases asymptotically towards $0.5$ in agreement with the predicted value for $c$. 
Near the transition the system shows remarkable finite-size effects 
around $\log \lambda \sim 5.5$, except for $U=5.396$ where data for different $L$ lies nearly on top of each other. We identify this as the Ising phase transition.

\subsection{Conclusion and Outlook}

We investigated the low-energy sector of the particle-hole subspace in the Bose-Hubbard model in the strong-coupling regime for $1D$ and $2D$. 
In first order $t/U$ we derived an effective two-particle problem perturbatively and revealed a free behavior in the (subspace) ground state independent of dimension and lattice.
We included a repulsive interacting term between nearest-neighbours and observed the lowering of the ground state and developing binding. The scaling behavior is independent 
of the lattice but depends strongly on the dimensionality. We derived second order terms and found an emergent binding behavior between the 
two quasi-particles up to this order. 
The bound state indeed has quantum numbers compatible with the CMI condensation and symmetry breaking in the intermediate regime, when the charge gap is small.

Using DMRG simulations on the 2-leg ladder with $\pi$-flux we measured a finite binding energy with a maximum at $U/t \sim 4.8$. We confirmed the 
exponentially closing of the particle-hole gap, prominent for the BKT universality class. By making a gap scaling ansatz we determined the thermodynamic limit 
for the transition $U_{\text{BKT}}/t = 4.10(1)$. We measured the exciton gap and identified the Ising transition with the closing of this transition at $U/t = 4.125(5)$.
 System sizes up to $L=512$ sites are too small to resolve this small CMI window ($\delta = 0.025(15)$) on the entropy level. 
We performed the same analysis on a zig-zag ladder and considered a finite repulsive nearest-neighbour interaction $V=1$ on both ladders. 
As the interaction immediately lowers the exciton energy in the strong coupling regime, we observed a CMI enlargement but by narrow margin.

Finally we studied the anisotropic coupling case $t_{\perp}/t=2$ at half-filling on the 2-leg ladder with $\pi$-flux and identified a very broad CMI phase with $\delta = 1.38(3)$.
In this case the Ising transition is clearly visible in the entanglement entropy and exhibits a prominent peak with corresponding free fermion central charge $c=1/2$.
In contrast to the unit-filling case, the CMI is sufficiently large to really measure it experimentally. It would be worthwhile exploring whether the 2d analogues of the rung Mott
insulators, the so called featureless Mott insulators~\cite{PhysRevLett.110.125301,featureless_Mott_1} 
can be used to engineer sizeable chiral Mott insulators in 2d.

\subsection*{Acknowledgments}
We thank M.~Dalmonte, F.~Heidrich-Meisner, A.~Sterdyniak and M.~Zalatel for discussions. We acknowledge support by the Austrian Science Fund for project SFB FoQus (F-4018). The computational results presented have been achieved in part using the Vienna Scientific Cluster (VSC). This work was supported by the Austrian Ministry of Science BMWF 
as part of the UniInfrastrukturprogramm of the Focal Point Scientific Computing at the University of Innsbruck. We are grateful to M.~Schuler and A.~Wietek for the usage of QuantiPy.

\bibliography{references_without_doi}

\end{document}